\documentclass[aps,superscriptaddress,prl,reprint]{revtex4-1}
\usepackage{ucs}
\usepackage[utf8x]{inputenc}
\usepackage{amsmath}
\usepackage{amsfonts}
\usepackage{amssymb}
\usepackage{amsthm}
\usepackage{fontenc}
\usepackage[dvips,pdftex]{graphicx}
\usepackage{setspace}
\usepackage{color}
\usepackage[dvipsnames]{xcolor}
\usepackage{graphicx}

\usepackage{hyperref}

\usepackage{multirow}
\usepackage{array}
\usepackage{subcaption}
\usepackage{pdfpages}
\usepackage{tikz}
\usepackage{multimedia}
\usepackage{float}
\usepackage{rotating}
\usepackage{epstopdf}

\makeatletter
\@addtoreset{chapter}{part}
\makeatother

\newcommand{\dt}{\,\text{d}}

\newcommand{\pd}{\partial}
\newcommand{\Tr}{\text{Tr}}

\graphicspath{{}}

\hypersetup{linktocpage}

\begin{document}
\title{Elasticity and Fluctuations of Frustrated Nano-Ribbons }
\date{\today}
\author{Doron \surname{Grossman}}
\email[]{doron.grossman@mail.huji.ac.il}
\affiliation{Racah Institute of Physics, Hebrew University, Jeruslaem 91904, Israel}
\author{Eran \surname{Sharon} }
\email[]{erans@mail.huji.ac.il }
\affiliation {Racah Institute of Physics, Hebrew University, Jeruslaem 91904, Israel}
\author{Haim \surname{Diamant}}
\email[]{hdiamant@tau.ac.il}
\affiliation{Raymond and Beverly Sackler School of Chemistry, Tel Aviv University, Tel Aviv 6997801, Israel}

\begin{abstract}
We derive a reduced quasi-one-dimensional theory of geometrically frustrated elastic ribbons. Expressed in terms of geometric properties alone, it applies to ribbons over a wide range of scales, allowing the study of their elastic equilibrium, as well as thermal fluctuations. We use the theory to account for the twisted-to-helical transition of ribbons with spontaneous negative curvature, and the effect of fluctuations on the corresponding critical exponents. The persistence length of such ribbons changes non-monotonically with  the ribbon's width,  dropping to zero at the transition. This and other statistical properties qualitatively differ from those of non-frustrated fluctuating filaments.
\end{abstract}

\maketitle






Slender structures appear on many scales in both natural and man-made systems. Examples vary from the tendrils and seedpods of plants \cite{Gerbode2012,Armon2011}, through man-made responsive gels and elastomers \cite{Sawa2011}, to nanoscale structures such as graphene  sheets \cite{Meyer2007} and biomolecular self-assemblies made of peptides \cite{Ziserman2011a}, lipids \cite{Thomas1999,Singh2003}, and proteins \cite{Lara2011}. Many of these systems (and others) are frustrated in the sense that, even when free of constraints, they contain residual stresses. On the nanoscale, such frustration is particularly likely, either due to a small mismatch between the molecular assemblies and 3D Euclidean geometry \cite{Oda1999}, or because of the accumulation of defects in crystalline sheets \cite{Shenoy2008,Koskinen2009}.

Present theories address the statistical-mechanical properties of compatible slender structures \cite{Nelson1989book,Bouchiat1998,Liverpool1998,Panyukov2000,Giomi2010}, and the elastic equilibrium of frustrated thin sheets (incompatible plate theory) \cite{Efrati2009a,Efrati2009,Efrati2011}. There is no general theory for the combination of the two, i.e., one which models the statistical mechanics of frustrates slender structures. As indicated by the work of Ghafouri and Bruinsma \cite{Ghafouri2005}, who modeled a specific case of a frustrated ribbon, the behavior of such systems is qualitatively different from that of ordinary semiflexible filaments.

In this Letter we derive a general theory for the elasticity and statistical mechanics of frustrated elastic ribbons. A 1D energy functional is derived from the 2D incompatible plate theory. It describes any ribbon, irrespective of whether or not its geometry is compatible with 3D Euclidean space. Motivated by recent measurements on self-assembled supramolecular structures \cite{Ziserman2011,Oda1999,Adamcik2010,Armon2014}, we proceed to apply the model to a specific system possessing negative spontaneous curvature.


In the formalism of incompatible sheets \cite{Efrati2009a}, a 2D membrane is fully
described by its metric $a$ and curvature tensor $b$. Both must satisfy the Gauss-Minardi-Patterson-Codazzi (GMPC) equations. Every membrane is equipped also with two intrinsic reference fields, $\bar{a}$ and $\bar{b}$, which in general may not satisfy the GMPC constraints. A surface configuration can usually comply with either $\bar{a}$ or $\bar{b}$, but not with both, giving rise to residual stresses. Thus, fluctuations around the equilibrium configurations of incompatible sheets should perturb the softer energy terms in the Hamiltonian not around their minimum, but about some point on a nontrivial energy landscape, giving rise to new physics.

The elastic 2D Hamiltonian  of a thin membrane is given by \cite{Efrati2009a,Armon2011}:
\begin{align}\label{eq:hamiltonian2D}
H=& \frac{Y }{8(1-\nu^2)}\iint\left(t E_s + \frac{t^3}{3}  E_b \right) \dt^2 A 
\\ \nonumber
E_s =&\nu \Tr^2[\bar{a}^{-1}(a-\bar{a})] + (1-\nu) \Tr[\bar{a}^{-1}(a-\bar{a})]^2  \\ \nonumber 
E_b =& \nu \Tr^2[\bar{a}^{-1}(b-\bar{b})] + (1-\nu) \Tr[\bar{a}^{-1}(b-\bar{b})]^2 ,
\end{align}
where $E_s$ is the stretching content, $E_b$ the bending content, $Y$ Young's modulus, $\nu$ Poisson's ratio, $t$ the thickness of the sheet (taken as the smallest length scale in the system), $\dt^2 A = \sqrt{\det{\bar{a}}}\; \dt u \dt v$ the surface element ($u$ and $v$ being a coordinate system on the sheet), $\bar{a}, \bar{b}$ the \emph{reference} metric and curvature, and $a,b$ the actual metric and curvature (the so-called first and second fundamental forms of a specific configuration). 

We consider a long (length $L$), narrow (width $W$) and thin (thickness $t$) ribbon, such that $ t \ll W \ll L $. We select a preferable set of coordinates, $(x,y) \in [0,L] \times [-\frac{W}{2},\frac{W}{2}]$, such that the mid-line of the ribbon is given by $(x,0)$. We reduce Eq. (\ref{eq:hamiltonian2D}) into a 1D Hamiltonian through an expansion of the curvatures around the mid-line in small $y$ (compared to the typical radius of curvature). The expansion is self-consistent in the sense that $a$ and $b$ describe a surface (i.e, the GMPC equations hold up to the required order). Only then do we allow the system to find the preferred configuration. 

 Near the mid-line, the reference metric may be approximated by
 $$ \bar{a}= \left( \begin{array}{cc}
			 (1+\bar{\kappa}_g y)^2 - \bar{K} y^2 + O(y^3) & 0 \\
			 0 & 1\\
			 \end{array} \right), $$			 
provided that 	$W < 2/\bar{\kappa}_g,~ 2/\sqrt{\bar{K}}$, where $\bar{\kappa}_g =\bar{\kappa}_g(x)$ is the \emph{reference} geodesic curvature of the mid-line, and $\bar{K} =\bar{K}(x)$  the \emph{reference} Gaussian curvature (i.e.,  if $a=\bar{a}$ these will coincide with the actual geodesic and Gaussian curvatures).
 In a similar manner we expand $\bar{b}$,
$$\bar{b} =  \left( \begin{array}{cc}
			\bar{l} & \bar{m} \\
			\bar{m} & \bar{n}\\
			\end{array} \right) + O(y). $$ 
As before, the elements of $\bar{b}$ are the curvatures at the mid-line (and may be a function of the position along the ribbon). The orders of these expansions for $\bar{a}$ and $\bar{b}$ are chosen such that a consistent expansion of the ultimate equations of equilibrium is obtained.

While the reference fields ($\bar{a},\bar{b}$) may not satisfy any particular relation, the actual metric and curvature tensors must satisfy the GMPC equations. A self-consistent expansion of these tensors gives (see supplementary material for details),

\begin{subequations}\label{eq:abfinal}
\begin{equation}\label{eq:a0f}
a = \left( \begin{array}{cc}
(1+\bar{\kappa}_g y)^2 - (l n - m^2) y^2   &    0\\
0& 1   \\ 
\end{array} 
\right) +  O(y^2),
\end{equation}
\begin{equation}\label{eq:b0f}
b = \left( \begin{array}{cc}
l + m'y & m + n'y \\
m + n'y & n\\
\end{array} \right) + O(y),
	\end{equation}
\end{subequations}
where we have defined $' \equiv \pd_x $. 
In addition, we set $\kappa_g = \bar{\kappa}_g$, since it is easily shown that deviations from this equality are too costly energetically.

\begin{figure}[tcb]
	\centering
	\includegraphics[clip=true,trim=0cm 0 0cm 0,height=0.2\textwidth,]{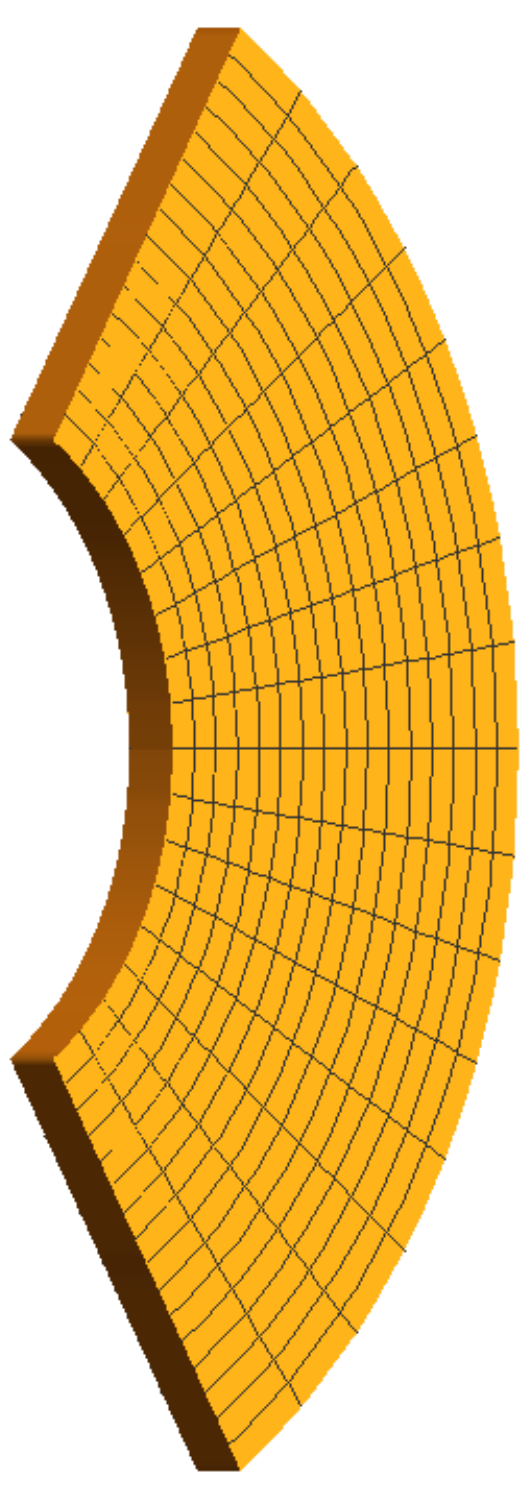}
	\hspace{1 em}
	\includegraphics[clip=true,trim=0cm 0 0cm 0,height=0.2\textwidth]{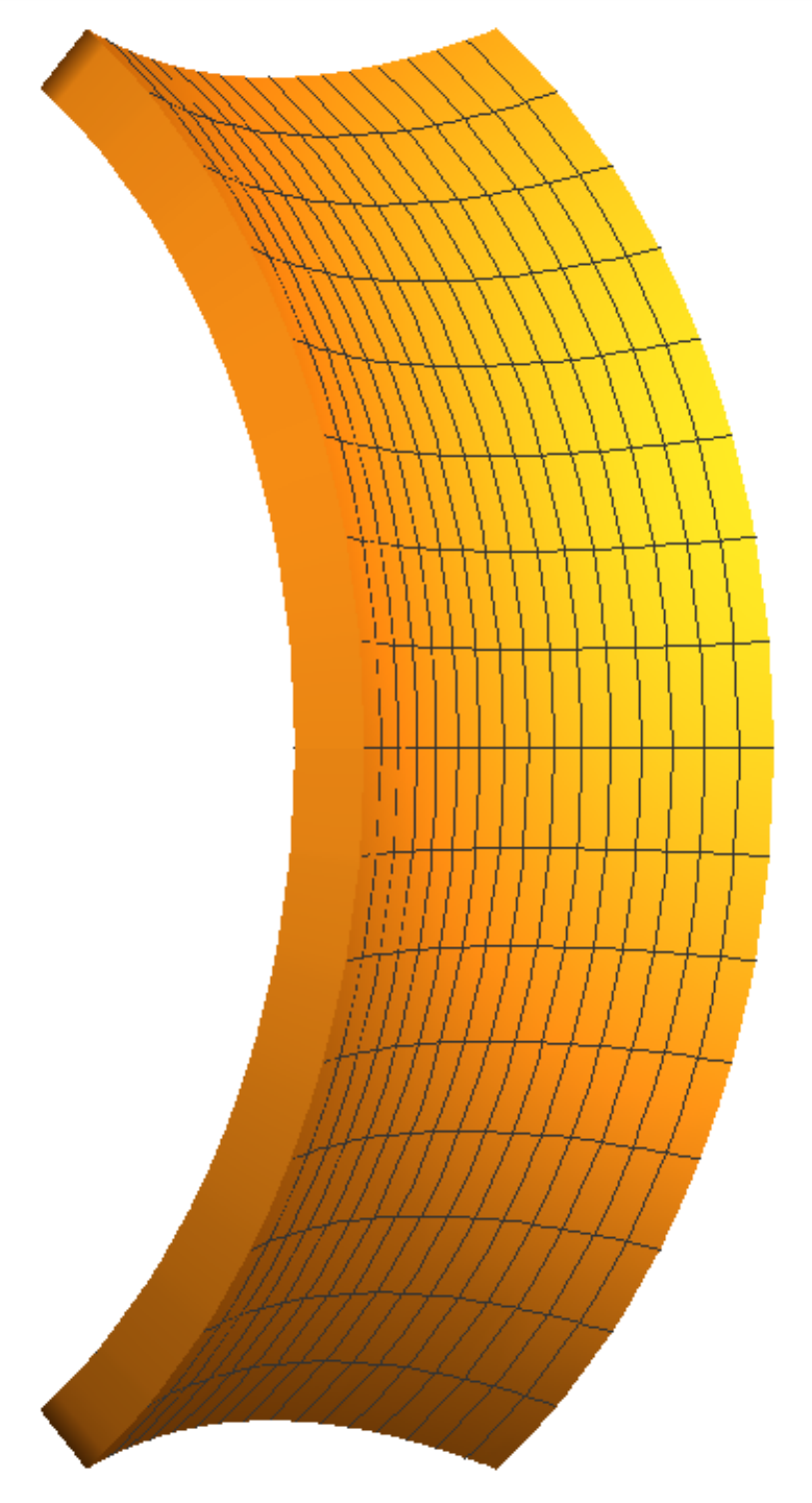}
	\hspace{.8 em}
	\includegraphics[clip=true,trim=0cm 0 0cm 0,height=0.2\textwidth]{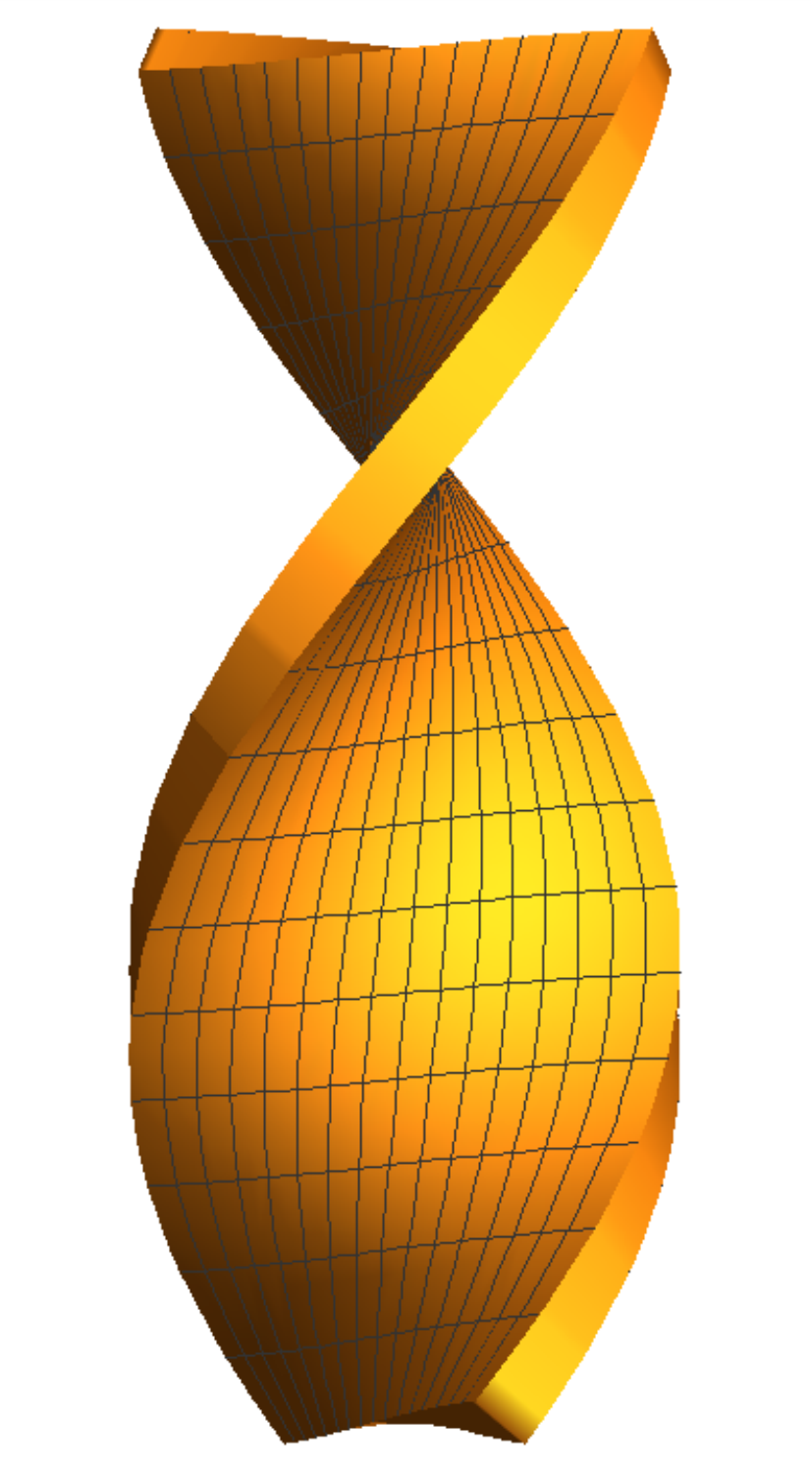}
 \label{fig:curvatures} 
\caption{\raggedright Geometric interpretation of the different curvatures. Left-to-right: $k_g$, $l$ and $n$ (opposite signs), $m$. \label{fig:curv_def}}
\end{figure}

Inserting the above expressions in the Hamiltonian (\ref{eq:hamiltonian2D}) and integrating over the narrow coordinate, $y$, we end up with

\begin{align}\label{eq:hamiltonian1D}
\nonumber
H =& \frac{Y}{8 \left(1-\nu ^2\right)} \int
\frac{1}{80} t W^5 \left(\bar{K}- l n + m^2\right)^2  
\\  &
+ \frac{1}{3} t^3 W \left[2 (1-\nu ) \biggl( \frac{W^2}{12} \left(n'\right)^2 \biggr. \right.  \\ \nonumber  & \left. \biggl.  
- \left( \left( \bar{l} - l\right) \left(\bar{n} - n \right)-\left(\bar{m} - m \right)^2 \right) \biggr) 
\right. \\ \nonumber & \left.
+ \left( \bar{l} + \bar{n}-n-l \right)^2 
+\frac{W^2}{12} \left(m'\right)^2
\right] \dt x.
\end{align}  
The first row in Eq. (\ref{eq:hamiltonian1D}) is the
stretching content, associated with deviation from $\bar{a}$. Other
rows are the bending content, related to the deviation from $\bar{b}$.

 Equation. (\ref{eq:hamiltonian1D}) is the central result of the present work, providing the energy functional for a wide range of compatible and incompatible ribbons. 
The quasi-1D reduction significantly simplifies the problem, allowing for analytical solution in simple cases (such as the one treated below), and the study of thermal fluctuations around the minimum. 

 A limit which is often studied is that of an unstretchable ribbon. This limit holds when the stretching rigidity is much larger than the bending rigidity, as in the limit $t \rightarrow 0$ .
This limit implies $a=\bar{a}$, i.e., the system is an isometric immersion (if it exists) of $\bar{a}$ in 3D Euclidean space.  If, in addition, we consider the case of a compatible, Euclidean ribbon  with no spontaneous curvature, i.e., $\bar{K}=\bar{\kappa}_g = \bar{l}=\bar{m}=\bar{n}\equiv 0$, we recover the known Sadowsky functional \cite{Sadowsky1930,Wunderlich1962},
 \begin{align} \label{eq:sadowsky} \nonumber
 H & =  \frac{Y}{8(1-\nu^2)} \int \frac{1}{3} t^3 W(l+n)^2  \dt x \\
 &=  \frac{Y t^3 W}{24(1-\nu^2)} \int \frac{(\kappa^2 + \tau^2)^2}{\kappa^2} \dt x, 
 \end{align}
The last equality was derived by solving for $n$ from the
relation (Gauss' theorema egregium) $ nl-m^2=0$ for a Euclidean membrane, and setting $l = \kappa$, $m = \tau$ in accordance with the Frenet-Serret frame.

We now use the Hamiltonian of Eq. (\ref{eq:hamiltonian1D}) to find an  analytic solution to the system described in Ref. \cite{Armon2014}. This is the case of a Euclidean reference metric,  $\bar{K}=0$, and spontaneous saddle curvature, $\bar{l}=\bar{n}=0, \bar{m}=k_0$. Such intrinsic geometry commonly appears in nano-ribbons generated by the self-assembly of chiral molecules. These include lipids \cite{Thomas1995}, peptides \cite{Ziserman2011a}, and proteins that form amyloids \cite{Adamcik2012}. The corresponding reference tensors are
\begin{equation}\label{eq:abarbbar}
\bar{a}= \left(\begin{array}{cc}
1 & 0\\
0 & 1
\end{array}\right), ~~~~~~~~~
\bar{b}= \left(\begin{array}{cc}
0 & k_0\\
k_0 & 0
\end{array}\right).
\end{equation} 
 The Hamiltonian of this specific system is then given by (omitting the derivatives in Eq. (\ref{eq:hamiltonian1D}), which are small in this case):
\begin{align}\label{eq:ebrs}
H =& \frac{Y}{8 \left(1-\nu ^2\right)} \int
\frac{1}{80} t W^5 \left(l n - m^2\right)^2
\\  \nonumber & 
+ \frac{1}{3} t^3 W \biggl[  \left( l+n\right)^2 
- 2 (1-\nu ) \left( l n -\left(k_0-m \right)^2 \right) \biggr]   \dt x.
\end{align}

\begin{figure}[tcb]
	\begin{subfigure}[h]{0.45\textwidth}
		\centering
	\begin{tikzpicture}
		        \node[anchor=south west,inner sep=0] (image) at (0,0) {\includegraphics[width=\textwidth]{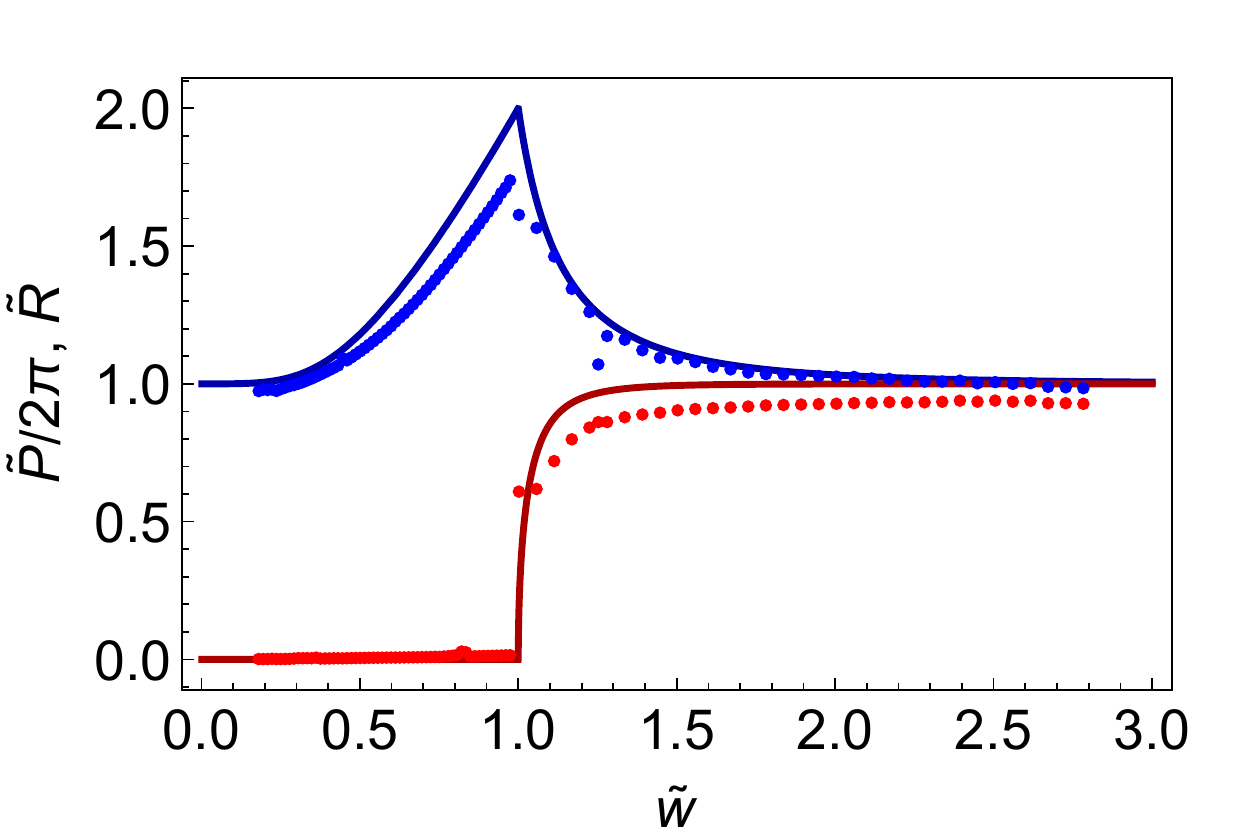}};
		    \end{tikzpicture}
	\end{subfigure}
	\caption{\raggedright Comparison between analytical model (solid lines), and 2D simulations (empty circles). Insets describe the ribbons configuration before (left) and after (right) the transition. \label{fig:comparison}}
\end{figure}

The minimum is found by solving the appropriate Euler-Lagrange  equations (see supplementary material). By defining dimensionless parameters: $\tilde{w} = W/W^*$, $\tilde{\sigma} =\sigma_i/k_0 ~ (\sigma_i \in \{l,m,n\})$, where  $ W^*= \left(\frac{320 (1+\nu)}{3 (1-\nu)^2} \frac{t^2}{k_0^2} \right)^{1/4}$, the solution gets the nifty form: 
\begin{subequations}\label{eq:dimless_sol}
	\begin{align}
		\tilde{n}=\tilde{l} =& \left\{ \begin{array}{c c}
			0 &  \tilde{w} \leq 1 \\
			\pm   \frac{1}{2} (1-\nu) \frac{\sqrt{\tilde{w}^4-1}}{ \tilde{w}^2} &   \tilde{w} >1
		\end{array} \right. \\
		\tilde{m}=& \left\{ \begin{array}{c c}
			\frac{1}{2} (1-\nu)  \ \frac{ \left( \tilde{\Xi}^2(\tilde{w}) -(1-\nu^2)3^{1/3} \right)}{ 3^{2/3} (1+ \nu )  \tilde{\Xi}(\tilde{w})  \tilde{w}^{2}  }&   \tilde{w} \leq 1\\
			\frac{1}{2}(1-\nu) &   \tilde{w} >1
		\end{array} \right. \\ \nonumber 
		\scriptstyle \tilde{\Xi}(\tilde{w})= & \scriptstyle \left[{\sqrt{   \left(81 (1+\nu )^4 \tilde{w}^4 + 3(1-\nu^2 )^3\right)}+9  (1+\nu)^2 \tilde{w}^2}\right]^{1/3}.  \\ \nonumber
	\end{align}
\end{subequations}
The dimensionless pitch and radius of the mid-line, $\tilde{P}= (1-\nu) k_0 \frac{ 2\pi m}{m^2 + l^2}$, $\tilde{R}= (1-\nu) k_0 \frac{ l}{m^2 + l^2}$,
can be compared with the solution obtained by 2D finite element numerics (Fig. \ref{fig:comparison}). Both  describe a  ribbon that changes its shape from  twisted to helical as it widens. 

Eqs. (\ref{eq:dimless_sol}) describe a continuous (second order) transition, at $W=W^*$ ($\tilde{w}=1$), which separates  two regimes: bending-dominated (i.e., minimization of the bending content is favored), and stretching-dominated (in which the solution of the bending-dominated regime becomes unstable). The $\pm$ signs mark a spontaneous symmetry breaking obtained by flipping a helix inside out (preserving chirality).  The twist to helical transition was observed in experiments and simulations \cite{Selinger2004,Armon2011,Sawa2011}, and the scaling (without numerical prefactors) of the critical width and existence of these two regimes were estimated in ref. \cite{Armon2011}. Qualitatively similar results were obtained within a modified 1D worm-like-chain model \cite{Ghafouri2005}.
The explicit solution, which we obtain from a controlled approximation of the 2D elastic problem, allows us to proceed and study in detail the statistical properties of the ribbons, on both sides of the transition.

 The nature of the elastic transition (Fig. \ref{fig:comparison}) is captured by its critical exponents, denoted by the conventional symbols $\alpha,\ \beta,\ \gamma,\ \delta,\ \nu_w,\ \eta$ \cite{chaikin2000principles}.
 The control parameter of the transition is the ribbon's width $\tilde{w}$.
Noting that the most singular behavior is observed for the mean curvature $\tilde{\Omega} \equiv\frac{1}{2}(\tilde{l} + \tilde{n})$, we immediately find the critical exponents  $\alpha = 0 ,~ \beta = \frac{1}{2},~ \gamma =1,~ \delta=3$. These results are presented in the supplementary material. 

We now turn to thermal fluctuations around the ground state, as these affect various observable properties of nanoribbons, such as the distributions of radius and pitch, and the persistence length. We expand the Hamiltonian (\ref{eq:hamiltonian1D}) around its equilibrium values to second order in the fluctuations $\Delta \sigma_i~ (\sigma_i \in \{l,m,n\})$, so that 
\begin{equation}\label{eq:2ndOrder}
H= H_{eq} + \int  \mathcal{H}^{(2)}_{i j} \Delta\sigma_i \Delta\sigma_j \dt x, 
\end{equation}
where $H^{(2)}_{i j} =  \frac{\pd^2 H}{\pd{\sigma_i} \pd{ \sigma_j}}  $ and we adopt the summation convention. Transforming to Fourier space ($x/W^* \rightarrow q$), and keeping only leading contributions near the transition, we obtain 
$ H-H_{eq} \propto \int (A_{\pm} |\delta \tilde{w}| + B_\pm q^2) | \Delta \tilde{\Omega}|^2 \dt q$ (See supplementary material for details), where $A_\pm,B_\pm$ are positive constants with different values below ($-$) and above ($+$) the transition. This calculation leads to the modified (but expected) values $\alpha= \frac{3}{2}$, $\nu_w = \frac{1}{2}$, $\eta=0$,
while $\beta$, $\gamma$, $\delta$ remain unchanged.
Due to the one dimensionality of the system ,the critical exponent $\alpha$ has  an atypical value. It is readily verified that the hyperscaling relation, $ 2-\alpha = \nu_w d$, is satisfied with $d=1$.

Calculating the fluctuations of the pitch and radius, within the Gaussian approximation, and integrating out $\Delta n$, we have

\begin{subequations}\label{eq:corr}
\begin{equation}
\langle \Delta \tilde{P}^2 \rangle = \frac{k_B T }{L Y k_0^{3/2} t^{7/2} } \tilde{\mathcal{H}}^{-1}_{pp} 
\end{equation}
\begin{equation}
\langle \Delta \tilde{R}^2 \rangle = \frac{k_B T }{L Y k_0^{3/2} t^{7/2} } \tilde{\mathcal{H}}^{-1}_{rr} 
\end{equation}
\begin{equation}
\langle \Delta \tilde{P}\Delta \tilde{R} \rangle = \frac{k_B T }{L Y k_0^{3/2} t^{7/2}}\tilde{\mathcal{H}}^{-1}_{rp} 
\end{equation}
\end{subequations}
where $\tilde{\mathcal{H}}$ is the dimensionless Hamiltonian. The functions $\tilde{\mathcal{H}}^{-1}_{ij}$ are shown in Fig. \ref{fig:corr} as a function of $\tilde{w}$. The  calculation and expressions are given in the supplementary material.

\begin{figure}[tcb]
	\centering
	\includegraphics[width=0.47\textwidth]{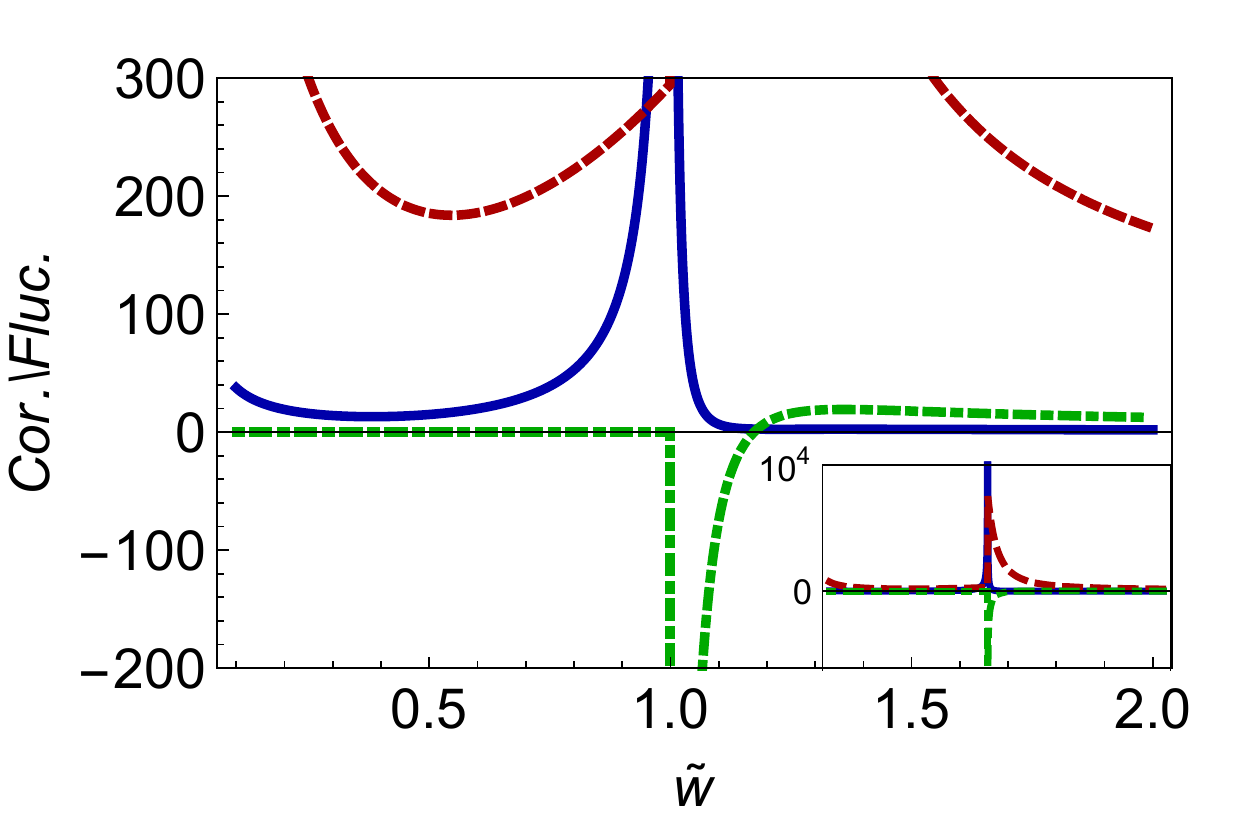}
	\caption{\raggedright Correlation and fluctuations as a function of (unitless) ribbon width, in the case of $\nu=0$. $\tilde{\mathcal{H}}^{-1}_{rr}$ in solid blue, $\tilde{\mathcal{H}}^{-1}_{rp}$ in dot-dashed green, $\tilde{\mathcal{H}}^{-1}_{pp}$ in dashed red. Inset: a `zoom out' view. Note that fluctuations in $\tilde{P}$ are finite (though discontinuous). For $\nu \neq 0$, features remain the same, though values vary. \label{fig:corr}}
\end{figure}

At the critical width, the fluctuation of the radius diverge due to the existence of an infinitely soft mode. Including the spatial ($q$) dependence of the fluctuations is important as it changes the divergence from  $|\tilde{w}-1|^{-1}$ ($q=0$ case)  to  $|\tilde{w}-1|^{-1/2}$. Right above the transition ($\tilde{w}>1$), the pitch and radius are negatively correlated (i.e., changes in the radius usually occur with an opposite change in the pitch). As the ribbon widens, the correlations change sign to become positive. This implies a change in the nature of the system's eigenmodes (and its response to small perturbations around the ground state) from over-winding to unwinding.

Knowing the fluctuations, we can calculate the persistence length, characterizing the bending fluctuations of the ribbon's mid-line, $l_p = \lim_{L \rightarrow \infty} \frac{1}{2L}\langle \vec{r}^2 \rangle$, where $\vec{r}$ is the ribbon's end-to-end vector. Intuitively, one expects the stiffness of the mid-line, and therefore $l_p$, to increase with the ribbons width, as indeed is found in non-frustrated ribbons \cite{Giomi2010}. Using the same formalism as in \cite{Panyukov2000a}, we find that the persistence length is non-monotonic in the width. Instead of continually increasing, it  drops to zero at the critical width. We find that near the transition, 
$$l_p = 	C_{\pm} \sqrt{|\tilde{w}-1|},$$
where $C_\pm$ are different positive constants below and above the transition.
Again, this result crucially depends on the integration over all $q$ of the fluctuations (yielding an exponent of $1/2$ instead of $1$).
Far above the transition, asymptotically, we have
$$l_p \xrightarrow{\tilde{w} \gg 1}\frac{2}{3}\frac{Y t^{3} }{k_B T} \frac{W^*}{(1-\nu^2)} \ \tilde{w}.$$  
This is a much larger result compared to non-frustrated ribbons, which is a direct consequence of the frustration.

\begin{figure}[tcb]
\centering
\includegraphics[width=0.47\textwidth]{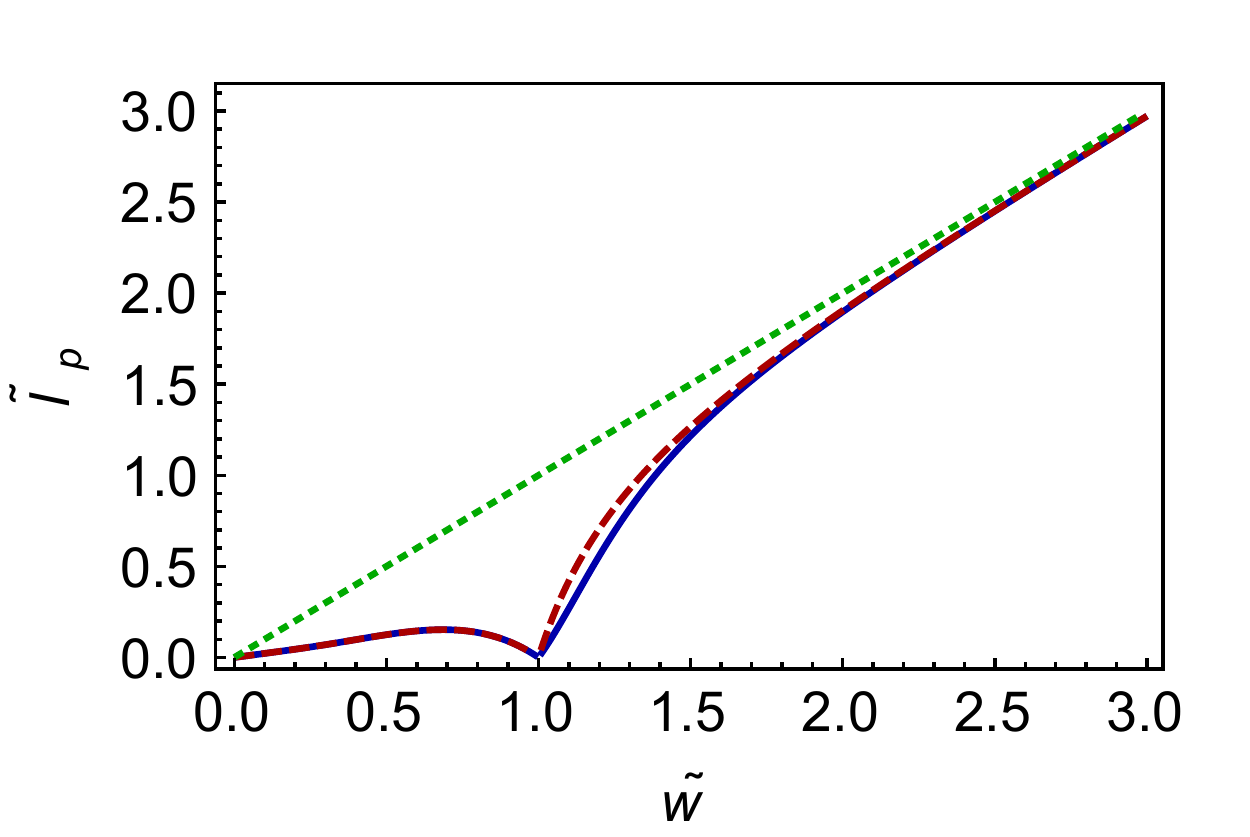}
\caption{\raggedright Dimensionless, normalized persistence length   $\tilde{l}_p = \frac{3(1-\nu^2)}{2} \frac{k_B T}{Y t^{3} W^*}  \ l_p. $ with $\frac{k_B T}{Y t^{3} W^*}=\psi=0.01$ (solid blue),$\psi=100$ (dashed red). Dotted green is the asymptote $\tilde{l}_p= \tilde{w}$ . \label{fig:persLenght}}
\end{figure}

We conclude with a few possible implications of our results for future experiments. The pitch and radius associated with the twist-to-helical transition (Fig. \ref{fig:comparison}) can be measured in nanometric self-assembled structures. Similar curves were found in macroscopic systems \cite{Armon2011}, while for the nanometric system of \cite{Ziserman2011a} it was only qualitatively characterized. Our work provides new results regarding the statistical nature of thermal systems --- specifically, the fluctuations in pitch and radius, and their correlations, and the persistence length of the ribbon's axis. The pitch and radius may be observed  using electron microscopies \cite{Ziserman2011,Oda1999}. The persistence length and its unusual dependence on the ribbon's width can be measured using light scattering. In particular, our theory predicts that, close to the transition, the persistence length should become very small. Under these conditions long ribbons should behave as random coils. The properties of a suspension of such ribbons should qualitatively change and resemble a solution of flexible polymers \cite{Rubinstein2003}.

We expect the results presented here, and in particular the reduced Hamiltonian of Eq. (\ref{eq:hamiltonian1D}), to be relevant to a range of biological, chemical, and condensed matter systems, in which fluctuating frustrated ribbons are known to exist.

\subsection{Acknowledgment}
The authors thank The Jacob Blaustein Institutes for Desert Research for their  hospitality. ES and DG  were supported by the European Research Council SoftGrowth project. DG was also supported by The Harvey M. Kruger Family Center of Nanoscience and Nanotechnology.

\bibliography{ribbons_arxiv}

\begin{thebibliography}{31}%
\makeatletter
\providecommand \@ifxundefined [1]{%
 \@ifx{#1\undefined}
}%
\providecommand \@ifnum [1]{%
 \ifnum #1\expandafter \@firstoftwo
 \else \expandafter \@secondoftwo
 \fi
}%
\providecommand \@ifx [1]{%
 \ifx #1\expandafter \@firstoftwo
 \else \expandafter \@secondoftwo
 \fi
}%
\providecommand \natexlab [1]{#1}%
\providecommand \enquote  [1]{``#1''}%
\providecommand \bibnamefont  [1]{#1}%
\providecommand \bibfnamefont [1]{#1}%
\providecommand \citenamefont [1]{#1}%
\providecommand \href@noop [0]{\@secondoftwo}%
\providecommand \href [0]{\begingroup \@sanitize@url \@href}%
\providecommand \@href[1]{\@@startlink{#1}\@@href}%
\providecommand \@@href[1]{\endgroup#1\@@endlink}%
\providecommand \@sanitize@url [0]{\catcode `\\12\catcode `\$12\catcode
  `\&12\catcode `\#12\catcode `\^12\catcode `\_12\catcode `\%12\relax}%
\providecommand \@@startlink[1]{}%
\providecommand \@@endlink[0]{}%
\providecommand \url  [0]{\begingroup\@sanitize@url \@url }%
\providecommand \@url [1]{\endgroup\@href {#1}{\urlprefix }}%
\providecommand \urlprefix  [0]{URL }%
\providecommand \Eprint [0]{\href }%
\providecommand \doibase [0]{http://dx.doi.org/}%
\providecommand \selectlanguage [0]{\@gobble}%
\providecommand \bibinfo  [0]{\@secondoftwo}%
\providecommand \bibfield  [0]{\@secondoftwo}%
\providecommand \translation [1]{[#1]}%
\providecommand \BibitemOpen [0]{}%
\providecommand \bibitemStop [0]{}%
\providecommand \bibitemNoStop [0]{.\EOS\space}%
\providecommand \EOS [0]{\spacefactor3000\relax}%
\providecommand \BibitemShut  [1]{\csname bibitem#1\endcsname}%
\let\auto@bib@innerbib\@empty
\bibitem [{\citenamefont {Gerbode}\ \emph {et~al.}(2012)\citenamefont
  {Gerbode}, \citenamefont {Puzey}, \citenamefont {McCormick},\ and\
  \citenamefont {Mahadevan}}]{Gerbode2012}%
  \BibitemOpen
  \bibfield  {author} {\bibinfo {author} {\bibfnamefont {S.~J.}\ \bibnamefont
  {Gerbode}}, \bibinfo {author} {\bibfnamefont {J.~R.}\ \bibnamefont {Puzey}},
  \bibinfo {author} {\bibfnamefont {A.~G.}\ \bibnamefont {McCormick}}, \ and\
  \bibinfo {author} {\bibfnamefont {L.}~\bibnamefont {Mahadevan}},\ }\href@noop
  {} {\bibfield  {journal} {\bibinfo  {journal} {Science}\ }\textbf {\bibinfo
  {volume} {337}},\ \bibinfo {pages} {1087} (\bibinfo {year}
  {2012})}\BibitemShut {NoStop}%
\bibitem [{\citenamefont {Armon}\ \emph {et~al.}(2011)\citenamefont {Armon},
  \citenamefont {Efrati}, \citenamefont {Kupferman},\ and\ \citenamefont
  {Sharon}}]{Armon2011}%
  \BibitemOpen
  \bibfield  {author} {\bibinfo {author} {\bibfnamefont {S.}~\bibnamefont
  {Armon}}, \bibinfo {author} {\bibfnamefont {E.}~\bibnamefont {Efrati}},
  \bibinfo {author} {\bibfnamefont {R.}~\bibnamefont {Kupferman}}, \ and\
  \bibinfo {author} {\bibfnamefont {E.}~\bibnamefont {Sharon}},\ }\href
  {\doibase 10.1126/science.1203874} {\bibfield  {journal} {\bibinfo  {journal}
  {Science}\ }\textbf {\bibinfo {volume} {333}},\ \bibinfo {pages} {1726}
  (\bibinfo {year} {2011})},\ \Eprint {http://arxiv.org/abs/80053151969}
  {arXiv:80053151969} \BibitemShut {NoStop}%
\bibitem [{\citenamefont {Sawa}\ \emph {et~al.}(2011)\citenamefont {Sawa},
  \citenamefont {Ye}, \citenamefont {Urayama}, \citenamefont {Takigawa},
  \citenamefont {Gimenez-Pinto}, \citenamefont {Selinger},\ and\ \citenamefont
  {Selinger}}]{Sawa2011}%
  \BibitemOpen
  \bibfield  {author} {\bibinfo {author} {\bibfnamefont {Y.}~\bibnamefont
  {Sawa}}, \bibinfo {author} {\bibfnamefont {F.}~\bibnamefont {Ye}}, \bibinfo
  {author} {\bibfnamefont {K.}~\bibnamefont {Urayama}}, \bibinfo {author}
  {\bibfnamefont {T.}~\bibnamefont {Takigawa}}, \bibinfo {author}
  {\bibfnamefont {V.}~\bibnamefont {Gimenez-Pinto}}, \bibinfo {author}
  {\bibfnamefont {R.~L.~B.}\ \bibnamefont {Selinger}}, \ and\ \bibinfo {author}
  {\bibfnamefont {J.~V.}\ \bibnamefont {Selinger}},\ }\href {\doibase
  10.1073/pnas.1017658108} {\bibfield  {journal} {\bibinfo  {journal}
  {Proceedings of the National Academy of Sciences}\ }\textbf {\bibinfo
  {volume} {108}},\ \bibinfo {pages} {6364} (\bibinfo {year} {2011})},\ \Eprint
  {http://arxiv.org/abs/http://www.pnas.org/content/108/16/6364.full.pdf}
  {http://www.pnas.org/content/108/16/6364.full.pdf} \BibitemShut {NoStop}%
\bibitem [{\citenamefont {Meyer}\ \emph {et~al.}(2007)\citenamefont {Meyer},
  \citenamefont {Geim}, \citenamefont {Katsnelson}, \citenamefont {Novoselov},
  \citenamefont {Booth},\ and\ \citenamefont {Roth}}]{Meyer2007}%
  \BibitemOpen
  \bibfield  {author} {\bibinfo {author} {\bibfnamefont {J.~C.}\ \bibnamefont
  {Meyer}}, \bibinfo {author} {\bibfnamefont {A.~K.}\ \bibnamefont {Geim}},
  \bibinfo {author} {\bibfnamefont {M.}~\bibnamefont {Katsnelson}}, \bibinfo
  {author} {\bibfnamefont {K.}~\bibnamefont {Novoselov}}, \bibinfo {author}
  {\bibfnamefont {T.}~\bibnamefont {Booth}}, \ and\ \bibinfo {author}
  {\bibfnamefont {S.}~\bibnamefont {Roth}},\ }\href@noop {} {\bibfield
  {journal} {\bibinfo  {journal} {Nature}\ }\textbf {\bibinfo {volume} {446}},\
  \bibinfo {pages} {60} (\bibinfo {year} {2007})}\BibitemShut {NoStop}%
\bibitem [{\citenamefont {Ziserman}\ \emph
  {et~al.}(2011{\natexlab{a}})\citenamefont {Ziserman}, \citenamefont {Lee},
  \citenamefont {Raghavan}, \citenamefont {Mor},\ and\ \citenamefont
  {Danino}}]{Ziserman2011a}%
  \BibitemOpen
  \bibfield  {author} {\bibinfo {author} {\bibfnamefont {L.}~\bibnamefont
  {Ziserman}}, \bibinfo {author} {\bibfnamefont {H.-Y.}\ \bibnamefont {Lee}},
  \bibinfo {author} {\bibfnamefont {S.~R.}\ \bibnamefont {Raghavan}}, \bibinfo
  {author} {\bibfnamefont {A.}~\bibnamefont {Mor}}, \ and\ \bibinfo {author}
  {\bibfnamefont {D.}~\bibnamefont {Danino}},\ }\href {\doibase
  10.1021/ja107069f} {\bibfield  {journal} {\bibinfo  {journal} {Journal of the
  American Chemical Society}\ }\textbf {\bibinfo {volume} {133}},\ \bibinfo
  {pages} {2511} (\bibinfo {year} {2011}{\natexlab{a}})}\BibitemShut {NoStop}%
\bibitem [{\citenamefont {Thomas}\ \emph {et~al.}(1999)\citenamefont {Thomas},
  \citenamefont {Lindemann},\ and\ \citenamefont {Clark}}]{Thomas1999}%
  \BibitemOpen
  \bibfield  {author} {\bibinfo {author} {\bibfnamefont {B.~N.}\ \bibnamefont
  {Thomas}}, \bibinfo {author} {\bibfnamefont {C.~M.}\ \bibnamefont
  {Lindemann}}, \ and\ \bibinfo {author} {\bibfnamefont {N.~A.}\ \bibnamefont
  {Clark}},\ }\href@noop {} {\bibfield  {journal} {\bibinfo  {journal}
  {Physical Review E}\ }\textbf {\bibinfo {volume} {59}},\ \bibinfo {pages}
  {3040} (\bibinfo {year} {1999})}\BibitemShut {NoStop}%
\bibitem [{\citenamefont {Singh}\ \emph {et~al.}(2003)\citenamefont {Singh},
  \citenamefont {Wong},\ and\ \citenamefont {Schnur}}]{Singh2003}%
  \BibitemOpen
  \bibfield  {author} {\bibinfo {author} {\bibfnamefont {A.}~\bibnamefont
  {Singh}}, \bibinfo {author} {\bibfnamefont {E.~M.}\ \bibnamefont {Wong}}, \
  and\ \bibinfo {author} {\bibfnamefont {J.~M.}\ \bibnamefont {Schnur}},\
  }\href {\doibase 10.1021/la026337r} {\bibfield  {journal} {\bibinfo
  {journal} {Langmuir}\ }\textbf {\bibinfo {volume} {19}},\ \bibinfo {pages}
  {1888} (\bibinfo {year} {2003})}\BibitemShut {NoStop}%
\bibitem [{\citenamefont {Lara}\ \emph {et~al.}(2011)\citenamefont {Lara},
  \citenamefont {Adamcik}, \citenamefont {Jordens},\ and\ \citenamefont
  {Mezzenga}}]{Lara2011}%
  \BibitemOpen
  \bibfield  {author} {\bibinfo {author} {\bibfnamefont {C.}~\bibnamefont
  {Lara}}, \bibinfo {author} {\bibfnamefont {J.}~\bibnamefont {Adamcik}},
  \bibinfo {author} {\bibfnamefont {S.}~\bibnamefont {Jordens}}, \ and\
  \bibinfo {author} {\bibfnamefont {R.}~\bibnamefont {Mezzenga}},\ }\href
  {\doibase 10.1021/bm200216u} {\bibfield  {journal} {\bibinfo  {journal}
  {Biomacromolecules}\ }\textbf {\bibinfo {volume} {12}},\ \bibinfo {pages}
  {1868} (\bibinfo {year} {2011})}\BibitemShut {NoStop}%
\bibitem [{\citenamefont {Oda}\ \emph {et~al.}(1999)\citenamefont {Oda},
  \citenamefont {Huc}, \citenamefont {Schmutz}, \citenamefont {Candau},\ and\
  \citenamefont {MacKintosh}}]{Oda1999}%
  \BibitemOpen
  \bibfield  {author} {\bibinfo {author} {\bibfnamefont {R.}~\bibnamefont
  {Oda}}, \bibinfo {author} {\bibfnamefont {I.}~\bibnamefont {Huc}}, \bibinfo
  {author} {\bibfnamefont {M.}~\bibnamefont {Schmutz}}, \bibinfo {author}
  {\bibfnamefont {S.~J.}\ \bibnamefont {Candau}}, \ and\ \bibinfo {author}
  {\bibfnamefont {F.~C.}\ \bibnamefont {MacKintosh}},\ }\href@noop {}
  {\bibfield  {journal} {\bibinfo  {journal} {Nature}\ }\textbf {\bibinfo
  {volume} {399}},\ \bibinfo {pages} {566} (\bibinfo {year}
  {1999})}\BibitemShut {NoStop}%
\bibitem [{\citenamefont {Shenoy}\ \emph {et~al.}(2008)\citenamefont {Shenoy},
  \citenamefont {Reddy}, \citenamefont {Ramasubramaniam},\ and\ \citenamefont
  {Zhang}}]{Shenoy2008}%
  \BibitemOpen
  \bibfield  {author} {\bibinfo {author} {\bibfnamefont {V.~B.}\ \bibnamefont
  {Shenoy}}, \bibinfo {author} {\bibfnamefont {C.~D.}\ \bibnamefont {Reddy}},
  \bibinfo {author} {\bibfnamefont {A.}~\bibnamefont {Ramasubramaniam}}, \ and\
  \bibinfo {author} {\bibfnamefont {Y.~W.}\ \bibnamefont {Zhang}},\ }\href
  {\doibase 10.1103/PhysRevLett.101.245501} {\bibfield  {journal} {\bibinfo
  {journal} {Physical review letters}\ }\textbf {\bibinfo {volume} {101}},\
  \bibinfo {pages} {245501} (\bibinfo {year} {2008})}\BibitemShut {NoStop}%
\bibitem [{\citenamefont {Koskinen}\ \emph {et~al.}(2009)\citenamefont
  {Koskinen}, \citenamefont {Malola},\ and\ \citenamefont
  {H{\"a}kkinen}}]{Koskinen2009}%
  \BibitemOpen
  \bibfield  {author} {\bibinfo {author} {\bibfnamefont {P.}~\bibnamefont
  {Koskinen}}, \bibinfo {author} {\bibfnamefont {S.}~\bibnamefont {Malola}}, \
  and\ \bibinfo {author} {\bibfnamefont {H.}~\bibnamefont {H{\"a}kkinen}},\
  }\href@noop {} {\bibfield  {journal} {\bibinfo  {journal} {Physical Review
  B}\ }\textbf {\bibinfo {volume} {80}},\ \bibinfo {pages} {073401} (\bibinfo
  {year} {2009})}\BibitemShut {NoStop}%
\bibitem [{\citenamefont {Nelson}\ \emph {et~al.}(1989)\citenamefont {Nelson},
  \citenamefont {Piran},\ and\ \citenamefont {Weinberg}}]{Nelson1989book}%
  \BibitemOpen
  \bibfield  {author} {\bibinfo {author} {\bibfnamefont {D.}~\bibnamefont
  {Nelson}}, \bibinfo {author} {\bibfnamefont {T.}~\bibnamefont {Piran}}, \
  and\ \bibinfo {author} {\bibfnamefont {S.}~\bibnamefont {Weinberg}},\
  }\href@noop {} {\emph {\bibinfo {title} {Statistical mechanics of membranes
  and surfaces, Volume 5 of Jerusalem Winter School for Theoretical Physics}}}\
  (\bibinfo  {publisher} {World Scientific, Singapore},\ \bibinfo {year}
  {1989})\BibitemShut {NoStop}%
\bibitem [{\citenamefont {Bouchiat}\ and\ \citenamefont
  {M\'ezard}(1998)}]{Bouchiat1998}%
  \BibitemOpen
  \bibfield  {author} {\bibinfo {author} {\bibfnamefont {C.}~\bibnamefont
  {Bouchiat}}\ and\ \bibinfo {author} {\bibfnamefont {M.}~\bibnamefont
  {M\'ezard}},\ }\href {\doibase 10.1103/PhysRevLett.80.1556} {\bibfield
  {journal} {\bibinfo  {journal} {Phys. Rev. Lett.}\ }\textbf {\bibinfo
  {volume} {80}},\ \bibinfo {pages} {1556} (\bibinfo {year}
  {1998})}\BibitemShut {NoStop}%
\bibitem [{\citenamefont {Liverpool}\ \emph {et~al.}(1998)\citenamefont
  {Liverpool}, \citenamefont {Golestanian},\ and\ \citenamefont
  {Kremer}}]{Liverpool1998}%
  \BibitemOpen
  \bibfield  {author} {\bibinfo {author} {\bibfnamefont {T.~B.}\ \bibnamefont
  {Liverpool}}, \bibinfo {author} {\bibfnamefont {R.}~\bibnamefont
  {Golestanian}}, \ and\ \bibinfo {author} {\bibfnamefont {K.}~\bibnamefont
  {Kremer}},\ }\href {\doibase 10.1103/PhysRevLett.80.405} {\bibfield
  {journal} {\bibinfo  {journal} {Phys. Rev. Lett.}\ }\textbf {\bibinfo
  {volume} {80}},\ \bibinfo {pages} {405} (\bibinfo {year} {1998})}\BibitemShut
  {NoStop}%
\bibitem [{\citenamefont {Panyukov}\ and\ \citenamefont
  {Rabin}(2000{\natexlab{a}})}]{Panyukov2000}%
  \BibitemOpen
  \bibfield  {author} {\bibinfo {author} {\bibfnamefont {S.}~\bibnamefont
  {Panyukov}}\ and\ \bibinfo {author} {\bibfnamefont {Y.}~\bibnamefont
  {Rabin}},\ }\href {\doibase 10.1103/PhysRevLett.85.2404} {\bibfield
  {journal} {\bibinfo  {journal} {Physical Review Letters}\ }\textbf {\bibinfo
  {volume} {85}},\ \bibinfo {pages} {2404} (\bibinfo {year}
  {2000}{\natexlab{a}})}\BibitemShut {NoStop}%
\bibitem [{\citenamefont {Giomi}\ and\ \citenamefont
  {Mahadevan}(2010)}]{Giomi2010}%
  \BibitemOpen
  \bibfield  {author} {\bibinfo {author} {\bibfnamefont {L.}~\bibnamefont
  {Giomi}}\ and\ \bibinfo {author} {\bibfnamefont {L.}~\bibnamefont
  {Mahadevan}},\ }\href {\doibase 10.1103/PhysRevLett.104.238104} {\bibfield
  {journal} {\bibinfo  {journal} {Physical Review Letters}\ }\textbf {\bibinfo
  {volume} {104}},\ \bibinfo {pages} {238104} (\bibinfo {year}
  {2010})}\BibitemShut {NoStop}%
\bibitem [{\citenamefont {Efrati}\ \emph
  {et~al.}(2009{\natexlab{a}})\citenamefont {Efrati}, \citenamefont {Sharon},\
  and\ \citenamefont {Kupferman}}]{Efrati2009a}%
  \BibitemOpen
  \bibfield  {author} {\bibinfo {author} {\bibfnamefont {E.}~\bibnamefont
  {Efrati}}, \bibinfo {author} {\bibfnamefont {E.}~\bibnamefont {Sharon}}, \
  and\ \bibinfo {author} {\bibfnamefont {R.}~\bibnamefont {Kupferman}},\ }\href
  {\doibase 10.1016/j.jmps.2008.12.004} {\bibfield  {journal} {\bibinfo
  {journal} {Journal of the Mechanics and Physics of Solids}\ }\textbf
  {\bibinfo {volume} {57}},\ \bibinfo {pages} {762} (\bibinfo {year}
  {2009}{\natexlab{a}})},\ \Eprint {http://arxiv.org/abs/0810.2411}
  {arXiv:0810.2411} \BibitemShut {NoStop}%
\bibitem [{\citenamefont {Efrati}\ \emph
  {et~al.}(2009{\natexlab{b}})\citenamefont {Efrati}, \citenamefont {Sharon},\
  and\ \citenamefont {Kupferman}}]{Efrati2009}%
  \BibitemOpen
  \bibfield  {author} {\bibinfo {author} {\bibfnamefont {E.}~\bibnamefont
  {Efrati}}, \bibinfo {author} {\bibfnamefont {E.}~\bibnamefont {Sharon}}, \
  and\ \bibinfo {author} {\bibfnamefont {R.}~\bibnamefont {Kupferman}},\ }\href
  {\doibase 10.1103/PhysRevE.80.016602} {\bibfield  {journal} {\bibinfo
  {journal} {Physical Review E}\ }\textbf {\bibinfo {volume} {80}},\ \bibinfo
  {pages} {016602} (\bibinfo {year} {2009}{\natexlab{b}})},\ \Eprint
  {http://arxiv.org/abs/arXiv:0902.2841v2} {arXiv:arXiv:0902.2841v2}
  \BibitemShut {NoStop}%
\bibitem [{\citenamefont {Efrati}\ \emph {et~al.}(2011)\citenamefont {Efrati},
  \citenamefont {Sharon},\ and\ \citenamefont {Kupferman}}]{Efrati2011}%
  \BibitemOpen
  \bibfield  {author} {\bibinfo {author} {\bibfnamefont {E.}~\bibnamefont
  {Efrati}}, \bibinfo {author} {\bibfnamefont {E.}~\bibnamefont {Sharon}}, \
  and\ \bibinfo {author} {\bibfnamefont {R.}~\bibnamefont {Kupferman}},\ }\href
  {\doibase 10.1103/PhysRevE.83.046602} {\bibfield  {journal} {\bibinfo
  {journal} {Physical Review E}\ }\textbf {\bibinfo {volume} {83}},\ \bibinfo
  {pages} {046602} (\bibinfo {year} {2011})}\BibitemShut {NoStop}%
\bibitem [{\citenamefont {Ghafouri}\ and\ \citenamefont
  {Bruinsma}(2005)}]{Ghafouri2005}%
  \BibitemOpen
  \bibfield  {author} {\bibinfo {author} {\bibfnamefont {R.}~\bibnamefont
  {Ghafouri}}\ and\ \bibinfo {author} {\bibfnamefont {R.}~\bibnamefont
  {Bruinsma}},\ }\href {\doibase 10.1103/PhysRevLett.94.138101} {\bibfield
  {journal} {\bibinfo  {journal} {Physical Review Letters}\ }\textbf {\bibinfo
  {volume} {94}},\ \bibinfo {pages} {138101} (\bibinfo {year}
  {2005})}\BibitemShut {NoStop}%
\bibitem [{\citenamefont {Ziserman}\ \emph
  {et~al.}(2011{\natexlab{b}})\citenamefont {Ziserman}, \citenamefont {Mor},
  \citenamefont {Harries},\ and\ \citenamefont {Danino}}]{Ziserman2011}%
  \BibitemOpen
  \bibfield  {author} {\bibinfo {author} {\bibfnamefont {L.}~\bibnamefont
  {Ziserman}}, \bibinfo {author} {\bibfnamefont {A.}~\bibnamefont {Mor}},
  \bibinfo {author} {\bibfnamefont {D.}~\bibnamefont {Harries}}, \ and\
  \bibinfo {author} {\bibfnamefont {D.}~\bibnamefont {Danino}},\ }\href
  {\doibase 10.1103/PhysRevLett.106.238105} {\bibfield  {journal} {\bibinfo
  {journal} {Physical Review Letters}\ }\textbf {\bibinfo {volume} {106}},\
  \bibinfo {pages} {238105} (\bibinfo {year} {2011}{\natexlab{b}})}\BibitemShut
  {NoStop}%
\bibitem [{\citenamefont {Adamcik}\ \emph {et~al.}(2010)\citenamefont
  {Adamcik}, \citenamefont {Jung}, \citenamefont {Flakowski}, \citenamefont
  {{De Los Rios}}, \citenamefont {Dietler},\ and\ \citenamefont
  {Mezzenga}}]{Adamcik2010}%
  \BibitemOpen
  \bibfield  {author} {\bibinfo {author} {\bibfnamefont {J.}~\bibnamefont
  {Adamcik}}, \bibinfo {author} {\bibfnamefont {J.-m.}\ \bibnamefont {Jung}},
  \bibinfo {author} {\bibfnamefont {J.}~\bibnamefont {Flakowski}}, \bibinfo
  {author} {\bibfnamefont {P.}~\bibnamefont {{De Los Rios}}}, \bibinfo {author}
  {\bibfnamefont {G.}~\bibnamefont {Dietler}}, \ and\ \bibinfo {author}
  {\bibfnamefont {R.}~\bibnamefont {Mezzenga}},\ }\href {\doibase
  10.1038/nnano.2010.59} {\bibfield  {journal} {\bibinfo  {journal} {Nature
  nanotechnology}\ }\textbf {\bibinfo {volume} {5}},\ \bibinfo {pages} {423}
  (\bibinfo {year} {2010})}\BibitemShut {NoStop}%
\bibitem [{\citenamefont {Armon}\ \emph {et~al.}(2014)\citenamefont {Armon},
  \citenamefont {Aharoni}, \citenamefont {Moshe},\ and\ \citenamefont
  {Sharon}}]{Armon2014}%
  \BibitemOpen
  \bibfield  {author} {\bibinfo {author} {\bibfnamefont {S.}~\bibnamefont
  {Armon}}, \bibinfo {author} {\bibfnamefont {H.}~\bibnamefont {Aharoni}},
  \bibinfo {author} {\bibfnamefont {M.}~\bibnamefont {Moshe}}, \ and\ \bibinfo
  {author} {\bibfnamefont {E.}~\bibnamefont {Sharon}},\ }\href {\doibase
  10.1039/c3sm52313f} {\bibfield  {journal} {\bibinfo  {journal} {Soft matter}\
  }\textbf {\bibinfo {volume} {10}},\ \bibinfo {pages} {2733} (\bibinfo {year}
  {2014})}\BibitemShut {NoStop}%
\bibitem [{\citenamefont {Sadowsky}(1930)}]{Sadowsky1930}%
  \BibitemOpen
  \bibfield  {author} {\bibinfo {author} {\bibfnamefont {M.}~\bibnamefont
  {Sadowsky}},\ }\href@noop {} {\bibfield  {journal} {\bibinfo  {journal}
  {Sitzungsber. Preuss. Akad. Wiss.}\ }\textbf {\bibinfo {volume} {22}},\
  \bibinfo {pages} {412–415} (\bibinfo {year} {1930})}\BibitemShut {NoStop}%
\bibitem [{\citenamefont {Wunderlich}(1962)}]{Wunderlich1962}%
  \BibitemOpen
  \bibfield  {author} {\bibinfo {author} {\bibfnamefont {W.}~\bibnamefont
  {Wunderlich}},\ }\href {\doibase 10.1007/BF01299052} {\bibfield  {journal}
  {\bibinfo  {journal} {Monatshefte für Mathematik}\ }\textbf {\bibinfo
  {volume} {66}},\ \bibinfo {pages} {276} (\bibinfo {year} {1962})}\BibitemShut
  {NoStop}%
\bibitem [{\citenamefont {Thomas}\ \emph {et~al.}(1995)\citenamefont {Thomas},
  \citenamefont {Safinya}, \citenamefont {Plano},\ and\ \citenamefont
  {Clark}}]{Thomas1995}%
  \BibitemOpen
  \bibfield  {author} {\bibinfo {author} {\bibfnamefont {B.~N.}\ \bibnamefont
  {Thomas}}, \bibinfo {author} {\bibfnamefont {C.~R.}\ \bibnamefont {Safinya}},
  \bibinfo {author} {\bibfnamefont {R.~J.}\ \bibnamefont {Plano}}, \ and\
  \bibinfo {author} {\bibfnamefont {N.~A.}\ \bibnamefont {Clark}},\ }\href
  {http://www.mrl.ucsb.edu/safinyagroup/PDFs/publications\_PDF/1995\_science\_1995\_267\_1635\_thomas\_safinya\_lipid\_tubule\_length.pdf}
  {\bibfield  {journal} {\bibinfo  {journal} {SCIENCE-NEW YORK \ldots}\ }
  (\bibinfo {year} {1995})}\BibitemShut {NoStop}%
\bibitem [{\citenamefont {Adamcik}\ and\ \citenamefont
  {Mezzenga}(2012)}]{Adamcik2012}%
  \BibitemOpen
  \bibfield  {author} {\bibinfo {author} {\bibfnamefont {J.}~\bibnamefont
  {Adamcik}}\ and\ \bibinfo {author} {\bibfnamefont {R.}~\bibnamefont
  {Mezzenga}},\ }\href {\doibase 10.1021/ma202157h} {\bibfield  {journal}
  {\bibinfo  {journal} {Macromolecules}\ }\textbf {\bibinfo {volume} {45}},\
  \bibinfo {pages} {1137} (\bibinfo {year} {2012})}\BibitemShut {NoStop}%
\bibitem [{\citenamefont {Selinger}\ \emph {et~al.}(2004)\citenamefont
  {Selinger}, \citenamefont {Selinger}, \citenamefont {Malanoski},\ and\
  \citenamefont {Schnur}}]{Selinger2004}%
  \BibitemOpen
  \bibfield  {author} {\bibinfo {author} {\bibfnamefont {R.}~\bibnamefont
  {Selinger}}, \bibinfo {author} {\bibfnamefont {J.}~\bibnamefont {Selinger}},
  \bibinfo {author} {\bibfnamefont {A.}~\bibnamefont {Malanoski}}, \ and\
  \bibinfo {author} {\bibfnamefont {J.}~\bibnamefont {Schnur}},\ }\href
  {\doibase 10.1103/PhysRevLett.93.158103} {\bibfield  {journal} {\bibinfo
  {journal} {Physical Review Letters}\ }\textbf {\bibinfo {volume} {93}},\
  \bibinfo {pages} {158103} (\bibinfo {year} {2004})}\BibitemShut {NoStop}%
\bibitem [{\citenamefont {Chaikin}\ and\ \citenamefont
  {Lubensky}(2000)}]{chaikin2000principles}%
  \BibitemOpen
  \bibfield  {author} {\bibinfo {author} {\bibfnamefont {P.~M.}\ \bibnamefont
  {Chaikin}}\ and\ \bibinfo {author} {\bibfnamefont {T.~C.}\ \bibnamefont
  {Lubensky}},\ }\href@noop {} {\emph {\bibinfo {title} {Principles of
  condensed matter physics}}},\ Vol.~\bibinfo {volume} {1}\ (\bibinfo
  {publisher} {Cambridge Univ Press},\ \bibinfo {year} {2000})\BibitemShut
  {NoStop}%
\bibitem [{\citenamefont {Panyukov}\ and\ \citenamefont
  {Rabin}(2000{\natexlab{b}})}]{Panyukov2000a}%
  \BibitemOpen
  \bibfield  {author} {\bibinfo {author} {\bibfnamefont {S.}~\bibnamefont
  {Panyukov}}\ and\ \bibinfo {author} {\bibfnamefont {Y.}~\bibnamefont
  {Rabin}},\ }\href {\doibase 10.1103/PhysRevE.62.7135} {\bibfield  {journal}
  {\bibinfo  {journal} {Physical Review E}\ }\textbf {\bibinfo {volume} {62}},\
  \bibinfo {pages} {7135} (\bibinfo {year} {2000}{\natexlab{b}})},\ \Eprint
  {http://arxiv.org/abs/0005317} {arXiv:0005317 [cond-mat]} \BibitemShut
  {NoStop}%
\bibitem [{\citenamefont {Rubinstein}\ and\ \citenamefont
  {Colby}(2003)}]{Rubinstein2003}%
  \BibitemOpen
  \bibfield  {author} {\bibinfo {author} {\bibfnamefont {M.}~\bibnamefont
  {Rubinstein}}\ and\ \bibinfo {author} {\bibfnamefont {R.~H.}\ \bibnamefont
  {Colby}},\ }\href@noop {} {\emph {\bibinfo {title} {Polymer physics}}}\
  (\bibinfo  {publisher} {OUP Oxford},\ \bibinfo {year} {2003})\BibitemShut
  {NoStop}%
\end{thebibliography}%


\begin{thebibliography}{2}%
\makeatletter
\providecommand \@ifxundefined [1]{%
 \@ifx{#1\undefined}
}%
\providecommand \@ifnum [1]{%
 \ifnum #1\expandafter \@firstoftwo
 \else \expandafter \@secondoftwo
 \fi
}%
\providecommand \@ifx [1]{%
 \ifx #1\expandafter \@firstoftwo
 \else \expandafter \@secondoftwo
 \fi
}%
\providecommand \natexlab [1]{#1}%
\providecommand \enquote  [1]{``#1''}%
\providecommand \bibnamefont  [1]{#1}%
\providecommand \bibfnamefont [1]{#1}%
\providecommand \citenamefont [1]{#1}%
\providecommand \href@noop [0]{\@secondoftwo}%
\providecommand \href [0]{\begingroup \@sanitize@url \@href}%
\providecommand \@href[1]{\@@startlink{#1}\@@href}%
\providecommand \@@href[1]{\endgroup#1\@@endlink}%
\providecommand \@sanitize@url [0]{\catcode `\\12\catcode `\$12\catcode
  `\&12\catcode `\#12\catcode `\^12\catcode `\_12\catcode `\%12\relax}%
\providecommand \@@startlink[1]{}%
\providecommand \@@endlink[0]{}%
\providecommand \url  [0]{\begingroup\@sanitize@url \@url }%
\providecommand \@url [1]{\endgroup\@href {#1}{\urlprefix }}%
\providecommand \urlprefix  [0]{URL }%
\providecommand \Eprint [0]{\href }%
\providecommand \doibase [0]{http://dx.doi.org/}%
\providecommand \selectlanguage [0]{\@gobble}%
\providecommand \bibinfo  [0]{\@secondoftwo}%
\providecommand \bibfield  [0]{\@secondoftwo}%
\providecommand \translation [1]{[#1]}%
\providecommand \BibitemOpen [0]{}%
\providecommand \bibitemStop [0]{}%
\providecommand \bibitemNoStop [0]{.\EOS\space}%
\providecommand \EOS [0]{\spacefactor3000\relax}%
\providecommand \BibitemShut  [1]{\csname bibitem#1\endcsname}%
\let\auto@bib@innerbib\@empty
\bibitem [{\citenamefont {Armon}\ \emph {et~al.}(2011)\citenamefont {Armon},
  \citenamefont {Efrati}, \citenamefont {Kupferman},\ and\ \citenamefont
  {Sharon}}]{Armon2011}%
  \BibitemOpen
  \bibfield  {author} {\bibinfo {author} {\bibfnamefont {S.}~\bibnamefont
  {Armon}}, \bibinfo {author} {\bibfnamefont {E.}~\bibnamefont {Efrati}},
  \bibinfo {author} {\bibfnamefont {R.}~\bibnamefont {Kupferman}}, \ and\
  \bibinfo {author} {\bibfnamefont {E.}~\bibnamefont {Sharon}},\ }\href
  {\doibase 10.1126/science.1203874} {\bibfield  {journal} {\bibinfo  {journal}
  {Science}\ }\textbf {\bibinfo {volume} {333}},\ \bibinfo {pages} {1726}
  (\bibinfo {year} {2011})},\ \Eprint {http://arxiv.org/abs/80053151969}
  {arXiv:80053151969} \BibitemShut {NoStop}%
\bibitem [{\citenamefont {Panyukov}\ and\ \citenamefont
  {Rabin}(2000)}]{Panyukov2000}%
  \BibitemOpen
  \bibfield  {author} {\bibinfo {author} {\bibfnamefont {S.}~\bibnamefont
  {Panyukov}}\ and\ \bibinfo {author} {\bibfnamefont {Y.}~\bibnamefont
  {Rabin}},\ }\href {\doibase 10.1103/PhysRevLett.85.2404} {\bibfield
  {journal} {\bibinfo  {journal} {Physical Review Letters}\ }\textbf {\bibinfo
  {volume} {85}},\ \bibinfo {pages} {2404} (\bibinfo {year}
  {2000})}\BibitemShut {NoStop}%
\end{thebibliography}%
\end{document}


\title{Supplementary Material- Quasi 1D Hamiltonian of frustrated ribbons}
\date{\today}
\author{Doron \surname{Grossman}}
\email[]{doron.grossman@mail.huji.ac.il}
\affiliation{Racah Institute of Physics, Hebrew University, Jeruslaem 91904, Israel}
\author{Eran \surname{Sharon} }
\email[]{erans@mail.huji.ac.il }
\affiliation {Racah Institute of Physics, Hebrew University, Jeruslaem 91904, Israel}
\author{Haim \surname{Diamant}}
\email[]{hdiamant@tau.ac.il}
\affiliation{Raymond and Beverly Sackler School of Chemistry, Tel Aviv University, Tel Aviv 6997801, Israel}


\maketitle

\appendix
\section{\label{app:quasi1D} Derivation of the Quasi-1D Hamiltonian}
This section includes a detailed derivation of Eq. (3) of the main article. We start with Eq. (1) and expand the curvature in $\bar{a}$ and $\bar{b}$ in powers of the narrow $|y|<W/2$ coordinate. We find that (to zeroth order)
\begin{equation*}
\bar{a}= \left( \begin{array}{cc}
			 (1+\bar{\kappa}_g y)^2 - \bar{K} y^2 + O(y^3) & 0 \\
			 0 & 1\\
			 \end{array} \right), $$$$
\bar{b} =  \left( \begin{array}{cc}
			\bar{l} & \bar{m} \\
			\bar{m} & \bar{n}\\
			\end{array} \right) + O(y),
\end{equation*}
where all the reference curvatures are estimated at the mid-line and may be functions of the $x$ coordinate. We now need to expand both $a$ and $b$ to the appropriate order. We first write:
\begin{equation*}
a \simeq  \left( \begin{array}{cc}
				(1+\kappa_g y)^2 - K y^2 &  \alpha y + \beta y^2 \\
				 \alpha y + \beta y^2 & 1 + \gamma y + \delta y^2
				\end{array}\right)  $$$$
b \simeq \left( \begin{array}{cc}
			l + l_{1} y & m + m_{1} y \\
			m + m_{1} y & n + n_{1} y
			\end{array}\right) 
\end{equation*}
where we have expanded $a$ to $2^{nd}$ order in a similar manner to what we have done for $\bar{a}$. The expansion of $b$ is straightforward. Next, we require self-consistency by solving the GMPC equations (in a perturbatively. These equations are given, to the leading order, by
\begin{subequations} 
			\begin{equation*}
				K + \frac{1}{2}\gamma \kappa_g + \alpha ' -(l n-m^2)=0 +O(y)
			\end{equation*}
			\begin{equation*}
				l_{1}=m' +\kappa_g(l + n) +O(y)
			\end{equation*}
			\begin{equation*}
				m_{1}=n' - \kappa_g m + \frac{1}{2} \gamma  m + \alpha l +O(y),
			\end{equation*}
\end{subequations}
where $'$ denotes a derivative along $x$. We can now calculate the Hamiltonian (by substituting these expressions for the fields, and integrating over $y$), and write the equations of motion. It can then be immediately shown that the solution for $\gamma,\ \alpha, \ \kappa_g$ is always (as long as $W \gg t$):
\begin{align*}
\kappa_g &= \bar{\kappa}_g \\
\alpha &= 0 \\
\gamma &= 0.
\end{align*}
This result is not surprising, as deviations of lower orders in $a-\bar{a}$ are energetically expensive. This can be seen by the following scaling argument (in principle similar to the one in Ref. \cite{Armon2011}).

Since we assume a frustrated ribbon (otherwise there is no problem to achieve $a=\bar{a}$ exactly), we compare the leading terms of the  stretching content (when $b\simeq\bar{b}$), and bending content (when $a\simeq \bar{a}$). Assuming that in a bending-dominated regime, the linear order of $a$ has some scale $\kappa$, while the linear order in $\bar{a}$ has a typical scale $\bar{\kappa}$, then the stretching content scales as $(\kappa-\bar{\kappa})^2 W^3 t$. On the other hand, in a stretching-dominated regime, deviation from $\bar{b}$ scales as $k-\bar{k}$, where $k$, $\bar{k}$ are the curvature scales of $b$, $\bar{b}$. Hence, the bending content scales as $(k-\bar{k})^2 t^3  W$. If we are to allow significant deviations from $\bar{a}$ we must have some regime such that the bending content dominates over the stretching content. One then requires- $(\kappa-\bar{\kappa})^2 W^3 t < t^3 (k-\bar{k})^2 W \Rightarrow  W  < t | \frac{k-\bar{k}}{\kappa-\bar{\kappa}}|$. As $W \gg t$, this can happen iff  either $\kappa$ is infinitely close to $\bar{\kappa}$, or $k$ is infinitely far from $\bar{k}$. Since the latter does not happen (otherwise our basic assumption that $t$ is the smallest scale in the system would be violated, hence Eq. (1)), we must conclude that the former is the only possibility. Thus, we may set $\kappa = \bar{\kappa}$ $\blacksquare$. We then find that the deviation from $\bar{a}$ (stretching content) actually scales, to leading order, as $W^5$, which gives rise to the same scaling argument as found in Ref. \cite{Armon2011}. It is worth noting that a significant deviation from $\bar{a}$ (i.e., including in zeroth and linear terms in $y$) may arise in cases of non-trivial boundary conditions.

We thus find that
\begin{equation*}
a \simeq  \left( \begin{array}{cc}
(1+\bar{\kappa}_g y)^2 - K y^2 &  \beta y^2 \\
\beta y^2 & 1 + \delta y^2
\end{array}\right),  $$$$
b \simeq \left( \begin{array}{cc}
l + l_{1} y & m + m_{1} y \\
m + m_{1} y & n + n_{1} y
\end{array}\right) ,
\end{equation*}
and the GMPC equations are of the form
\begin{subequations} \label{eq:GMPC}
	\begin{equation}\label{eq:egregium}
	K =(l n - m^2) +O(y)
	\end{equation}
	\begin{equation}\label{eq:gmpc1}
	l_{1}=m' +\kappa_g(l + n) +O(y)
	\end{equation}
	\begin{equation}\label{eq:gmpc2}
	m_{1}=n' - \kappa_g m +O(y).
	\end{equation}
\end{subequations}
We notice that Eqs. (\ref{eq:gmpc1}) and (\ref{eq:gmpc2}) include the next order correction of $b$. However, we note that derivatives along $x$ may diverge (e.g.  when there is a discontinuity in $\bar{b}$); hence, this part must be included in our expansion of $b$. We also note that within the required accuracy ($0^{th}$ order in the curvatures) the quadratic terms in $a$, except the one in $a_{11}$, do not contribute, and hence may be omitted. Finally, we conclude (using Eqs. (\ref{eq:GMPC})),

\begin{equation}
a \simeq  \left( \begin{array}{cc}
(1+\bar{\kappa}_g y)^2 - (l n-m^2) y^2 &  0 \\
0 & 1 
\end{array}\right),  $$$$
b \simeq \left( \begin{array}{cc}
l + m' y & m + n' y \\
m + n' y & n 
\end{array}\right) ,
\end{equation}
as in Eqs. (2a) and (2b) of the main text.

\section{ \label{app:critexp} Critical Exponents }
In this section we calculate explicitly the critical exponents of the systems. For simplicity, calculations are given for the case of vanishing Poisson ratio, $\nu=0$; the critical behavior for $\nu \neq 0$ remains the same. We start by calculating $\gamma$ and $\delta$. To this end we must introduce an external field. Since the mean curvature $\Omega$ is our order parameter, one can insert an external field as $\bar{l}$, $\bar{n}$, or a combination of both. We notice that the Hamiltonian (Eq. (1)) (and hence also Eq.(3)), is symmetric under rotation of the principal axes of curvature in $\bar{b}$. In the case of our 1D model, the mechanical and thermodynamic properties remain invariant under rotation, though the form of the solution may change. Hence, each $\bar{b}$ of the form  $$ \left( \begin{array}{cc}
 \bar{l} & k_0\\
   k_0  & \bar{n}
\end{array} \right) $$ with small enough $\bar{l}$ and $\bar{n}$ may be rotated to get the form of $$ \left( \begin{array}{cc}
 0 & k_0 \\
   k_0 & \alpha k_0
\end{array} \right), $$  where we have assumed that $\alpha$ is arbitrarily small, and hence the off-diagonal elements ($\bar{m}$) remain essentially the same. It is therefore sufficient to consider this case only. It is worth mentioning  that an exact solution to the problem just defined exists, although it is quite cumbersome. We will restrict the discussion here to the simplest approximation required  to evaluate the critical exponents.

We now substitute the above expression into our model, and find small corrections to the Euler-Lagrange equations. That is, we write $\sigma_i \equiv \sigma_{i,0} + \delta \sigma_i ,\ (\sigma_i \in \{l,m,n\})$ and expand the E.L. equations to leading orders, assuming (since $\alpha$ is small) $\delta \sigma_i$'s are small too.  We also define $$\delta \Omega = \frac{1}{2} (\delta l + \delta n),$$ $$\delta \theta = \frac{1}{2} (\delta l - \delta n),$$  $$\delta \tilde{w} = \tilde{w} -1,$$ and linearize the equations in $\delta \tilde{w}$. At the critical width $\delta \tilde{w}=0$ we have
\begin{subequations}
\begin{equation}
-2 \alpha -16 \delta\theta ^2 \delta\Omega -16 \delta m^2 \delta\Omega -16 \delta m \delta\Omega +16 \delta\Omega ^3=0
\end{equation}	
\begin{align}\nonumber
2 \alpha  & +16 \delta\theta ^3 +16 \delta\theta  \delta m^2+\\ &16 \delta\theta  \delta m-16 \delta\theta  \delta\Omega ^2+8 \delta\theta=0
\end{align}	
\begin{align}\nonumber
16 \delta\theta ^2 \delta m & +8 \delta\theta ^2-16 \delta m \delta\Omega ^2 +  \\ &  16 \delta m^3+24 \delta m^2+16 \delta m-8 \delta\Omega ^2=0.
\end{align}	
\end{subequations} As all parameters are assumed small we may write the above equations by keeping only leading orders

\begin{subequations} \label{eq:el_crit}
\begin{equation} \label{eq:el_crit1}
-2 \alpha+16 \delta\Omega ^3=0
\end{equation}	
\begin{equation}\label{eq:el_crit2}
2 \alpha +8 \delta\theta=0
\end{equation}	
\begin{equation}\label{eq:el_crit3}
16 \delta m=0,
\end{equation}	
\end{subequations} 
where we have used the fact the Eqs.0 (\ref{eq:el_crit2}) and (\ref{eq:el_crit3}) contain a linear term to simplify Eq. (\ref{eq:el_crit1}). From  this one immediately sees that the exponent $\delta =3$.

By taking the derivatives (with respect to $\alpha$)  of the linearized E.L. equations near the transition, and approximating them at $\alpha =0$, we find that $\chi = \pd_\alpha \Omega|_{\alpha =0}$ satisfies, near the transition,
\begin{equation}
\chi =\frac{2}{48 \delta\Omega^2  - 16 \delta \tilde{w}},
\end{equation}
where we have kept only terms up to linear order in $\delta \tilde{w}$, and used the fact that near the transition, $\delta{m}$ is either $0$ (above) or $\sim \delta \tilde{w}$. Thus, near the transition,
\begin{equation}
\chi \sim \left\{ \begin{array}{cc}
1/(16 \delta \tilde{w} )& \delta \tilde{w}>0 \\
-1/(8 \delta \tilde{w} )& \delta \tilde{w}<0 \\
\end{array} \right.,
\end{equation}
i.e., the exponent $\gamma =1$.

We include spatial derivatives. The Hamiltonian is given by
\begin{align}
H =& \frac{Y}{8 \left(1-\nu ^2\right)} \int
\frac{1}{80} t W^5 \left(l n - m^2\right)^2
+ \frac{1}{3} t^3 W \biggl[  \left( l + n\right)^2 
+ \frac{W^2}{12} m'^2 \biggr.    \biggl. \biggl.  - 2 (1-\nu ) \left(l n -\left(k_0-m\right)^2  - \frac{W^2}{12} n'^2\right)  \biggr]   \dt x.
\end{align}
We compute the partition function-
\begin{align}
\mathcal{Z} &\simeq \mathcal{Z}_{eq} \int \mathcal{D}[\{
\Delta\sigma_i \}] e^{- \beta {H}^{(2)}[{\Delta \sigma_i}]}, \\
\end{align}
where  $\beta =1/k_B T$,
\begin{align}
\nonumber
H^{(2)} =& \frac{Y}{8 \left(1-\nu ^2\right)} \int
\frac{1}{80} t W^5  \Bigl( \left( -\Delta n l -\Delta l n + \right. \Bigr. \Bigl. \left. 
 2 \Delta m m \right)^2  + 2 \left(\Delta m^2- \Delta n \Delta l \right)\left( m^2-l n \right) \Bigr)
\\  & 
+ \frac{1}{3} t^3 W \biggl[ \frac{W^2}{12} (\Delta m')^2 + (\Delta n +\Delta l)^2 + \biggr.  \biggl.
2(1-\nu)\left(\Delta m^2 - \Delta l \Delta n + \frac{W^2}{12} (\Delta n')^2 \right)  \biggr]   \dt x.
\end{align}
is the second order expansion of the Hamiltonian around the equilibrium solution, and the $\sigma_i$'s are given in Eqs. (7).

Since the Hamiltonian includes derivatives, there arises the need to use Fourier transformation. This requires some discussion relating the effective dimensionality of the system. It is clear, physically, that at high enough wavenumbers, the effective 1D description of the system is not valid anymore. At wavenumbers of order $O(1/W)$, the one-dimensionality of the system breaks down and one must treat it as  a 2D system (allowing fluctuations along the $y$ direction). 

As most of the singular behavior occurs at small wavenumbers we therefore limit the sum to $q \lesssim 1/W$, keeping in mind that there is another, nonsingular, part. By changing into dimensionless variables ($qW^* \rightarrow q$) we find the Hamiltonian
\begin{align}
\nonumber
H^{(2)} =& \frac{Y k_0^{3/2} t^{7/2}}{8 \left(1-\nu ^2\right)} \int\limits_{-1}^1
\frac{1}{80}  \tilde{w}^5  \Bigl( \left| -\Delta\tilde{n} \tilde{l} -\Delta\tilde{l} \tilde{n} + \right. \Bigr. \Bigl. \left. 
 2 \Delta \tilde{m} \tilde{m} \right|^2  + 2 \left(|\Delta\tilde{m}|^2- \Re( \Delta\tilde{n}\Delta\tilde{l}^\dagger)\right)\left(\tilde{m}^2-\tilde{l}\tilde{n}\right) \Bigr)
\\  & 
+ \frac{1}{3} \tilde{w} \biggl[ \frac{\tilde{w}^2}{12} q^2 |\Delta\tilde{m}|^2 + |\Delta \tilde{n} +\Delta \tilde{l}|^2 + \biggr. \biggl.
2(1-\nu)\left(|\Delta \tilde{m}|^2 - \Re(\Delta\tilde{l}\Delta\tilde{n}^\dagger) + \frac{\tilde{w}^2}{12} q^2 |\Delta\tilde{n}|^2 \right)  \biggr]   \dt q,
\end{align}
where $^\dagger$ marks complex conjugation.

For simplicity, we integrate out $\Delta\tilde{n}$. This results in an effective Hamiltonian, with the same critical behavior as the original one, in which  $\Delta \tilde{l}$ are the  fluctuations of the order parameter (and not $\Delta \tilde{l} +\Delta \tilde{n}$). It is easily verified that no singular behavior is added or removed (as the coefficient of $|\Delta\tilde{n}|^2$ is non-singular). We now write $H^{(2)}$ in a matrix form (over $\Delta \tilde{l}$ and $\Delta \tilde{m}$ near the transition $\tilde{w} =1$).  
$$H^{(2)} = Y k_0^{3/2} t^{7/2} \frac{(\frac{5}{3})^{1/4}}{3 \sqrt{2}} \int \mathcal{H}_{i j} \Delta \tilde{\sigma}_i \Delta \tilde{\sigma}_j  \dt q,$$ where $ i,j \in \{2,3\}$, and find that below the critical width ($\delta\tilde{w}<0$):
\begin{subequations}
\begin{align}
\mathcal{H}_{ll}&= \frac{q^2}{q^2+6} -\frac{\left(144+ 6 q^2 - q^4\right)}{\left(q^2+6\right)^2} \delta\tilde{w} \\
\mathcal{H}_{lm}&= 0 \\
\mathcal{H}_{mm}&= \frac{\left(q^2+96\right)}{12}+\frac{\left(q^2+80\right)}{4} \delta\tilde{w}  
\end{align}
\end{subequations}
 and above the transition ($\delta\tilde{w}>0$),
\begin{subequations}
\begin{align}
\mathcal{H}_{ll}&= \frac{q^2}{q^2+6}+\frac{\left(576 +114 q^2+ 5 q^4\right)}{\left(q^2+6\right)^2} \delta\tilde{w} \\
\mathcal{H}_{lm}&=  -\frac{4 \left(q^2+12\right)}{q^2+6} \sqrt{\delta\tilde{w}}  \\
\mathcal{H}_{mm}&= \frac{1}{12} \left(q^2+96\right)+\frac{ \left(192+102 q^2+q^4\right)}{4 \left(q^2+6\right)} \delta\tilde{w}.
\end{align}
\end{subequations}
It is immediately seen that the singular behavior comes solely from the $\mathcal{H}_{ll}$ terms.

We calculate  $\mathcal{H}^{-1}$ (as both correlations and the partition function are essentially functions of it), using the fact that $|q|<1$ (so that we may discard orders higher than quadratic). Keeping only singular contributions, we find that below the transition
\begin{subequations}
\begin{align}
(\mathcal{H}^{-1})_{ll}&= \frac{6}{ q^2-24 \ \delta\tilde{w} } \\
(\mathcal{H}^{-1})_{lm}&= 0 \\
(\mathcal{H}^{-1})_{mm}&= \frac{1}{ 8+ q^2/12 +20 \ \delta\tilde{w}},
\end{align}
\end{subequations}
 and above it
 \begin{subequations}
 \begin{align}
 (\mathcal{H}^{-1})_{ll}&= \frac{6}{q^2 +48 \ \delta\tilde{w} } \\
(\mathcal{H}^{-1})_{lm}&= \frac{6 \sqrt{\delta\tilde{w} }}{q^2 + 48 \ \delta\tilde{w}}\\
 (\mathcal{H}^{-1})_{mm}&= \frac{q^2+ 96 \delta\tilde{w}}{8 q^2+ 384 \ \delta \tilde{w}}.
 \end{align}
 \end{subequations}
 Notice that the most singular behavior comes from the $(\mathcal{H}^{-1})_{ll}$ element. Hence, it is sufficient to study its contribution alone. 
 
The correlations of the order parameter are given by
\begin{align}
\langle \Delta \tilde{l} (q) \Delta \tilde{l} (q') \rangle \simeq \langle \Delta \tilde{l} ^2 \rangle   \delta(q+q'), \\ \label{eq:corrfunfourier}
\langle \Delta \tilde{l} ^2 \rangle  \simeq \frac{k_B T}{Y k_0^{3/2} t^{7/2}} \ \frac{3 \sqrt{2}}{(\frac{5}{3})^{1/4}}\  \frac{6}{q^2 +\xi^{-2}},
\end{align}
where we have defined the dimensionless correlation length to be
\begin{align}\label{eq:corrlength}
\xi = \xi_{\pm} /\sqrt{|\delta\tilde{w}|},
\end{align}
and $\xi_{\pm}$ is either $1/48$ or $1/24$ depending on the sign of $\delta\tilde{w}$. Hence, it is clear that $\nu_w = 1/2$.

The spatial correlation is given by
\begin{align}
\langle \Delta\tilde{l}(0) \Delta\tilde{l}(r)  \rangle = \frac{1}{2\pi} \int\limits_{-1}^1 \langle \Delta \tilde{l} ^2 \rangle e^{-i q r }   \dt q + ...
\end{align}
where $r$ is the dimensionless distance between two points on the ribbon's mid-line. Given the expression (\ref{eq:corrfunfourier}), we can easily calculate this integral near the transition by defining $\mu= \xi q$:
 \begin{align*}
\langle \Delta\tilde{l}(0) \Delta\tilde{l}(r)  \rangle \propto \xi \int\limits_{-\xi}^\xi \frac{e^{-i \mu r/\xi}}{\mu^2 +1}  \dt \mu +... 
 \end{align*}
 Close to the transition we may take the bounds of integration to $\pm \infty$ (indicating that near the transition, the systems may indeed be treated as 1D), so that
 \begin{align*}
 \langle \Delta\tilde{l}(0) \Delta\tilde{l}(r)  \rangle \propto \frac{k_B T}{Y k_0^{3/2} t^{7/2} \sqrt{|\delta\tilde{w}|}} e^{-\frac{|r|}{\xi}},
 \end{align*}
 where we have used (\ref{eq:corrlength}). 
 This result (though strictly divergent at the critical point)  means $\eta =0$.
 
 Finally, calculating the contribution to the ``specific heat" is straightforward. Near the transition, the most singular eigenvalue of $\mathcal{H}^{-1}$ is  $(\mathcal{H}^{-1})_{22}$;  hence we find the dimensionless free energy,
 \begin{equation}
 \beta F \propto - \frac{1}{2} \int \log \left[\frac{k_B T}{Y k_0^{3/2} t^{7/2}} \ \frac{6}{q^2 +\xi_{\pm}^{-2} |\delta\tilde{w}|} \right] \dt q .
 \end{equation}
 Using the relation $C \propto - \pd^2_{\delta\tilde{w}} F$ we find that 
 \begin{equation}
C \propto  \int \frac{1}{(q^2 + \xi^{-2})^2} \dt q  \propto \xi^3 \int\limits_{\infty}^\infty \frac{1}{(\mu^2 +1)^2} \dt \mu $$$$
C \propto  \frac{1}{|\delta \tilde{w}|^{3/2}},
 \end{equation}
thus $\alpha=3/2$ .

\section{ \label{app:Fluc}  Dimensionless Correlation/Fluctuation function }
In this appendix we give results for the case of vanishing Poisson ration, $\nu=0$, If it is finite, the critical behavior remains the same, while the numerical factors may change somewhat. As in Appendix B, we first integrate out $\Delta\tilde{n}$. Subsequently, we change variables from $\Delta \tilde{l}$, $\Delta \tilde{m}$, to $\Delta{\tilde{R}}$, $\Delta\tilde{P}$. Above the transition ($\tilde{w}>1$) we find
\begin{subequations}
\begin{align}
\mathcal{H}^{-1}_{rr}& = \Psi \frac{\tilde{w}^4 \left(2 \tilde{w}^8-6 \tilde{w}^4+5\right)}{\tilde{w}^4-1} \\
\mathcal{H}^{-1}_{rp}& =\Psi \frac{\pi  \tilde{w}^2 \left(4 \tilde{w}^8-6 \tilde{w}^4-3\right)}{\sqrt{\tilde{w}^4-1}}\\
\mathcal{H}^{-1}_{pp}& =2 \Psi  \pi ^2 \left(8 \tilde{w}^8+4 \tilde{w}^4+1\right),
\end{align}
\end{subequations}
where
$$ \Psi= \frac{3 \sqrt{2}}{(\frac{5}{3})^{1/4}} \frac{8 \tilde{w}^7}{\left(1-2 \tilde{w}^4\right)^4} $$
Below the transition $(\tilde{w}<1)$
\begin{subequations}
\begin{align}
\mathcal{H}^{-1}_{rr}& =\frac{4 \times 3^{5/3} \sqrt{2}}{\sqrt[4]{\frac{5}{3}}} \tilde{w}^{15} \bigg/ \Biggl[ 3^{5/6} \Xi ^{2/3} \Sigma + \left(9 \sqrt[3]{\Xi }+8\ 3^{2/3}\right) \tilde{w}^{12}  -\sqrt{3} \sqrt[3]{\Xi } \Sigma  \tilde{w}^4 + \left(10 \sqrt[3]{3} \sqrt[3]{\Xi }+3\right) \sqrt[3]{\Xi } \left(-\tilde{w}^8\right) \Biggr] \\
\mathcal{H}^{-1}_{rp}& =0 \\
\mathcal{H}^{-1}_{rr}& =\frac{32 \times 9 \times 3^{5/3} \sqrt{2}}{\sqrt[4]{\frac{5}{3}}} \pi ^2 \Xi ^2 \tilde{w}^{19} \bigg/  \Biggl[\biggl( \sqrt[6]{3} \sqrt[3]{\Xi } \Sigma+ \biggr.\Biggr. \left(3\ 3^{2/3} \sqrt[3]{\Xi }+\sqrt[3]{3}\right) \tilde{w}^8 - \tilde{w}^4 \Xi ^{2/3} \biggr) \left(\Xi ^{2/3}-\sqrt[3]{3} \tilde{w}^4\right)^4 \Biggr],
\end{align}
\end{subequations}
where 
$$ \Sigma = \sqrt{27 y^{16}+y^{12}}, $$$$ \Xi =\sqrt{3} \Sigma +9 y^8.$$

\section{Calculating the Persistence Length}
Calculation of the persistence length is done with a slight generalization of the formalism shown in Ref.  \cite{Panyukov2000} to non-diagonalized matrices.

We start from the definition of the persistence length,
\begin{equation}\label{eq:perslength}
l_p= \lim_{L\rightarrow \infty} \frac{1}{2L}\langle \vec{r}^2 \rangle = \lim_{L\rightarrow \infty} \frac{1}{L}\int\limits_0^L \dt s \int\limits_0^s \dt s' \langle \vec{t}(s) \cdot \vec{t}(s') \rangle,
\end{equation}
where $\vec{r}$ is the end-to-end vector, $\vec{t}(s)$ is a unit vector tangent to the ribbon along the mid-line, $\langle X \rangle$ is the thermodynamic average of $X$, $s$ is the distance along the mid-line, and $L$ is the length of the ribbon.

If the ground state configuration has constant curvatures along the mid-line, then

\begin{equation}
\langle \vec{t}(s) \vec{t}(s') \rangle= \left[ e^{-\Lambda (s-s')}\right]_{33},
\end{equation}
where
\begin{equation}
\Lambda = \Omega + \frac{k_B T}{2} \left( I_{2\times 2} \Tr(\mathcal{H}_{(2)}^{-1}) - \mathcal{H}_{(2)}^{-1}\right).
\end{equation}
Here $ I_{2\times 2}$ is the $2\times2$ unit matrix,
$$\mathcal{H}_{(2)}= \left( \begin{array}{ccc}
\mathcal{H}_{ll} & \mathcal{H}_{l\kappa_g} & \mathcal{H}_{lm}\\
\mathcal{H}_{\kappa_g l} & \mathcal{H}_{\kappa_g \kappa_g} & \mathcal{H}_{\kappa_g m}\\
\mathcal{H}_{ml} & \mathcal{H}_{m \kappa_g} & \mathcal{H}_{mm}\\
\end{array}\right)$$ is the bending moduli matrix (the coefficients in the 2nd-order expansion of the 1D Hamiltonian in fluctuations around its ground state). Notice that in our case $\mathcal{H}_{i \kappa_g}$ is very large, and we may take it as infinite).
$$\Omega = \left( \begin{array}{ccc}
 0 & m & -\kappa_g \\
 -m & 0 & l \\
\kappa_g & -l  & 0 \\
\end{array} \right)
$$ is the infinitesimal rotation matrix, where we have identified the curvatures with Kirchoff's rotation vector formalism. 

\bibliography{ribbons_arxiv}